\documentclass{article}

\usepackage{arxiv}

\usepackage[utf8]{inputenc} 
\usepackage[T1]{fontenc}    
\usepackage{hyperref}       
\usepackage{url}            
\usepackage{booktabs}       
\usepackage{amsfonts}       
\usepackage{nicefrac}       
\usepackage{microtype}      
\usepackage{lipsum}
\usepackage{graphicx,verbatim,float,subfigure,graphicx,color,siunitx}
\usepackage{amsmath}
\graphicspath{ {./images/} }

\usepackage{amssymb}
\usepackage{svg}
\usepackage{multirow}  
\usepackage[export]{adjustbox}
\usepackage{soul}
\usepackage{textcomp}
\usepackage{multirow}
\usepackage{enumitem}
\usepackage{amsfonts}
\usepackage{bbm}
\usepackage{amsthm}

\title{Role of Material Directionality on the Mechanical Response of Miura-Ori Composite Structures} 

\author{
 Haotian Feng$^{\ast}$ \\
  Dept. of Mechanical Engineering\\
  University of Wisconsin-Madison \\
  Madison, WI 53706 \\
  \And
 Guanjin Yan\thanks{The authors have equal contributions.} \\
  Dept. of Civil \& Env. Engineering \\
  University of Wisconsin-Madison \\
  Madison, WI 53706 \\
  \And
  Pavana Prabhakar\thanks{Corresponding author.}\\
  Dept. of Mechanical Engineering \\
  Dept. of Civil \& Env. Engineering \\
  University of Wisconsin-Madison \\
  Madison, WI 53706 \\
  \texttt{pavana.prabhakar@wisc.edu} \\
}

\begin{document}
\maketitle

\begin{abstract}

This paper aims to understand the role of directional material properties on the mechanical responses of origami structures. We consider the Miura-Ori structures our target model due to their collapsibility and negative Poisson’s ratio (NPR) effects, which are widely used in shock absorbers, disaster shelters, aerospace applications, etc. Traditional Miura-Ori structures are made of isotropic materials (Aluminum, Acrylic), whose mechanical properties like stiffness and NPR are well understood. However, how these responses are affected by directional materials, like Carbon Fiber Reinforced Polymer (CFRP) composites, needs more in-depth understanding. To that end, we study how fiber directions and arrangements in CFRP composites and Miura-Ori's geometric parameters control the stiffness and NPR of such structures. Through finite element analysis, we show that Miura-Ori structures made of CFRP composites can achieve higher stiffness and Poisson’s ratio values than those made of an isotropic material like Aluminum. Then through regression analysis, we establish the relationship between different geometric parameters and the corresponding mechanical responses, which is further utilized to discover the Miura-Ori structure's optimal shape. We also show that the shear modulus is a dominant parameter that controls the mechanical responses mentioned above among the individual composite material properties within the Miura-Ori structure. We demonstrate that we can optimize the Miura-Ori structure by finding geometric and material parameters that result in combined stiffest and most compressible structures. We anticipate our research to be a starting point for designing and optimizing more sophisticated origami structures with composite materials incorporated.

\end{abstract}






\section{Introduction}\label{intro}

Deployable structures can change their shape such that their size can be significantly altered. This is often achieved by the folding and unfolding process, which provides deployable structures the flexibility to expand or contract based on different geometrical and material properties. Deployable structures are mostly used for easy storage and transport and are deployed into their operational configuration when required\cite{Pellegrino2001}. They find applications in astrophysics missions\cite{PUIG201012}, vehicle fairings\cite{reiman2004deployable}, and spacecrafts\cite{doi:10.2514/1.22854} among others. These structures often require to be lightweight, have high stiffness, and occupy less storage space. Among different ways of achieving deployable structures, origami structures, the ancient art of folding paper, have been extended to engineering applications to design deployable structures. 

Miura fold\cite{schenk2013geometry}, a classical origami pattern, is a method of folding a flat surface such as a sheet of paper into a smaller effective area. Miura-Ori patterns are composed of identical unit cells of mountain and valley folds with four-coordinated ridges. Different fold patterns could lead to interesting mechanical responses and material properties, like higher stiffness and contraction ability known as negative Poisson's ratio (NPR)\cite{lakes1987foam}. Materials such as foams and microporous polymers with NPR are often referred to as auxetics\cite{Caddock_1989}. They often possess the high energy-absorbing capability and fracture resistance. Miura-Ori patterns with such unique characteristics have been used in applications at both micro and macro scales, like in solar panels\cite{jasim2018origami}, surgical stents\cite{johnson2017fabricating}, and self-deployable robots\cite{mintchev2015foldable}. The origami or Miura-Ori patterns have also been widely used as impact mitigating systems\cite{yasuda2019origami} and impact protectors like the crush box\cite{yuan2019quasi} and robotic rotorcraft protectors\cite{sareh2018rotorigami}. Besides these structural applications, recent researchers also utilized the unique shape of Miura-Ori in other fields, including stretchable circuit bands\cite{li2021miura}, sound absorption\cite{wang2021ultralight}, aerodynamic drag reduction\cite{zhang2022deployment}, and stiffness vibration isolators\cite{ye2022origami}.

Past research has focused on exploring the mechanical responses of origami structures. Fisher et al.\cite{fischer2009sandwich} explored the stress-strain relationship (through compression and transverse shear tests) of sandwich structures with different types of cores made by folding sheet materials into three-dimensional zigzag patterns. Heimbs et al.\cite{heimbs2007experimental} investigated both experimental and numerical methods for analyzing folded cores called Ventable Shear Core (VeSCo) in sandwich composites under dynamic compression loading. With regard to Miura-Ori structures, Zhou et al.\cite{zhou2014mechanical} presented a parametric study on the mechanical responses of a variety of Miura-Ori-based folded core models by virtual testing under quasi-static compression, shear, and bending using the Finite Element Analysis (FEA). Wei et al.\cite{wei2013geometric} characterized the geometry and analyzed the effective elastic response of a simple periodically folded Miura-Ori structure by establishing mathematical expressions for stiffness and Poisson's ratio concerning geometric parameters. Liu et al.\cite{liu2015deformation} performed FEA based study to analyze the deformations of the Miura-Ori patterned sheet and validated the simulation result through tests under the same loading. Moradweysi et al.\cite{moradweysi2022design} analyzed the effective properties of Miura-Ori structures with small or significantly large thicknesses. The authors focused on analyzing unit cells of the Miura-Ori pattern under static periodic boundary conditions. They discovered that when the thickness is small, the rigid rotation of facets becomes dominant, and as the thickness value increases, the strict rotation mechanism tends to diminish. Gao et al.\cite{gao2022origami}proposed a novel design of Miura-Ori honeycomb structure, whose core layer at the middle is sandwiched by two secondary flange layers at the top and bottom. The model has connectivity with an open channel within the material and a self-locking feature under loading. Zhang et al.\cite{ZHANG2018411} proposed a thick panel Miura-Ori structure that uses bistable anti-symmetric carbon fiber reinforced polymer (CFRP) shells to connect and drive the whole structure. The authors proposed an energy-based theoretical model and showed that the proposed structure has superior stability, high reliability, and fast response speed. Besides these works, other researchers also look into the potentialities to design and optimize deployable structures with Machine Learning and Statistical Learning\cite{zhang2022space,zhu2022harnessing}.

These previous studies on folded structures, especially Miura-Ori structures, have focused on investigating the relationship between geometric parameters and mechanical responses for isotropic materials. However, few researchers have focused on combining composite materials and origami patterns to harness the benefits of directional properties and high stiffness/strength-to-weight ratios, like in fiber-reinforced composites. Domber et al.\cite{domber2002dimensional} studied the dimensional repeatability of elastically folded composite hinges for deployed spacecraft optics. They investigated a new type of folded composite hinge for precision deployable spacecraft structures and showed that viscoelastic recovery is independent of stow duration. Saito et al.\cite{saito2014manufacture} focused on manufacturing composite honeycomb cores based on origami patterns. They illustrated a new strategy to construct arbitrary cross-section composite honeycombs by applying 3D kirigami patterns into honeycombs and generalizing the honeycomb model by corresponding kirigami parameters. The authors further fabricated the kirigami honeycombs using the proposed folding line diagrams (FLD) design method. Cui et al.\cite{cui2018origami} investigated origami pattern guided morphing for composite sheets, where they demonstrated a new approach to fold origami/kirigami structures based on Gaussian curvature change induced by nonuniform lateral shrinkage and embedded rigid origami skeleton in polymer sheets to create novel 3D structures. Kwon et al.\cite{Hyundoi:10.1177/1045389X19873429} analyzed origami-inspired shape memory dual-matrix composite structures. They developed an analytical model to analyze 3D morphing structures and fabricated woven fabrics based on shape memory polymers. They further showed that morphing structures could be highly flexible depending on temperature based on tensile tests.

However, research on how structures made of composite materials with directional properties and geometric parameters of origami patterns influence their mechanical responses is lacking. Thus, in the current paper, we focus on exploring how the directionality of fiber-reinforced polymer composites and geometric parameters of the patterns impact the optimal shape of the Miura-Ori structure. To evaluate this, we consider mechanical properties - compressive stiffness and Poisson's ratio. We then determine the directional Miura-Ori structure that outperforms the isotropic Miura-Ori structure. 

In this paper, we focus on the following three objectives:

\begin{enumerate}

\item {\bf Understand how fiber directionality influences the mechanical responses of Miura-Ori structures.} 

For the directional material, we consider the transversely isotropic material (Carbon Fiber Reinforced Polymer composites - CFRP composites\cite{hyer2009stress}). We vary the geometry and fiber arrangements within the Miura-Ori cells. We then perform compression analysis using the finite element method. We use Miura-Ori cells with isotropic material, Aluminium, as a baseline model to compare against the mechanical properties of different CFRP composite arrangements. We consider CFRP composites due to their inherent transversely isotropic material behavior. CFRP composites are extremely strong and lightweight compared to traditional isotropic materials and are widely used in engineering, including aerospace, automotive, marine, and construction\cite{feng2021difference}.

\item {\bf Explore how geometric parameters affect Miura-Ori's corresponding mechanical response and determine the optimal design}

We fix the isotropic and directional materials and change the geometric parameters to find the optimal design. Specifically, we perform a regression analysis to determine the relationship between geometric parameters and the corresponding mechanical responses. We then use this relationship to determine the optimal geometric configuration subject to different optimal requirements. 

\item {\bf Establish how individual material properties affect the overall mechanical responses of Miura-Ori structures}

After the geometric analysis of Miura-Ori, we vary individual material properties of composite materials to elucidate what properties have a dominant effect on the mechanical responses, like stiffness and NPR. We propose using woven composites for potentially altering a few specific material properties due to the flexibility of selecting different yarn materials and weaving patterns\mbox{\cite{penava2015woven,naik1992elastic,feng2022physics}}.

\end{enumerate}

Although the results shown in this paper are for the Miura-Ori structure, the same methodologies in this paper can be extended to the analysis and optimization of other types of origami structures.

\section{Miura-Ori Model Setup}
\subsection{Model Geometry and Directionality}

\begin{figure}[h!]
\centering
\subfigure[]{
	\includegraphics[width=0.29\textwidth]{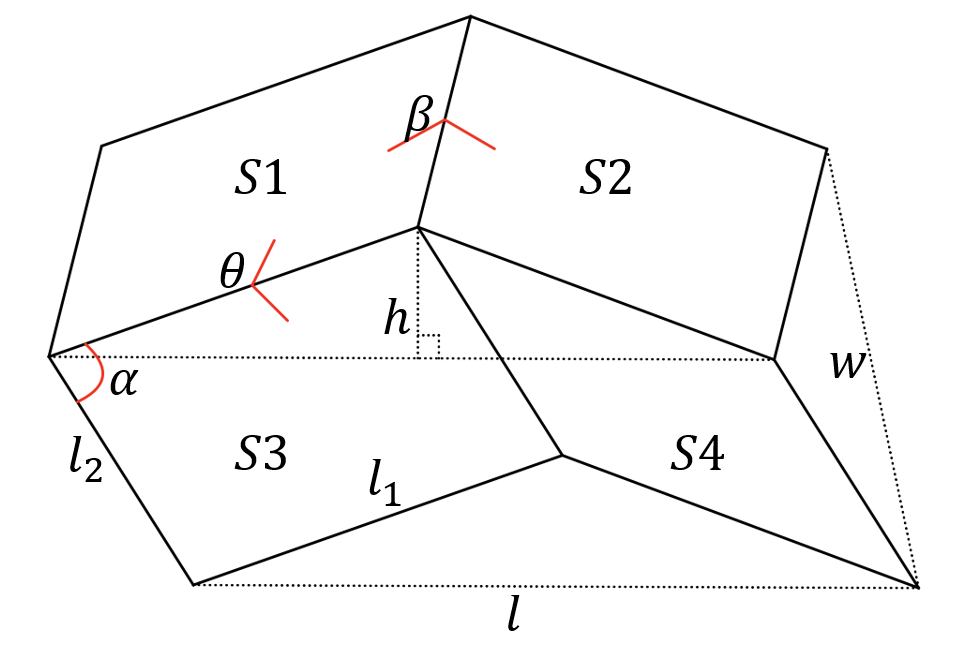}
	}
\centering
\subfigure[]{
	\includegraphics[width=0.31\textwidth]{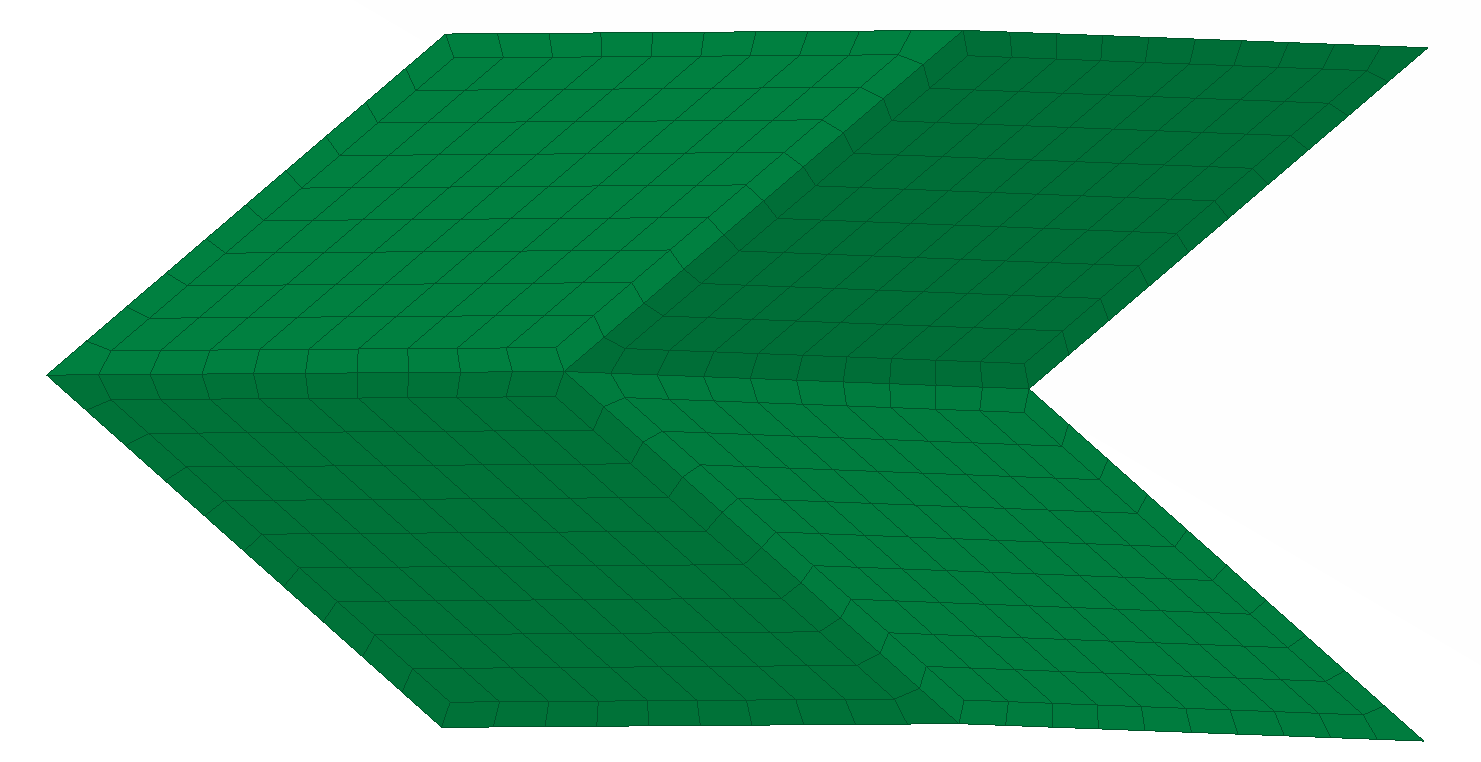}
	}
\subfigure[]{
    \includegraphics[width=0.31\textwidth]{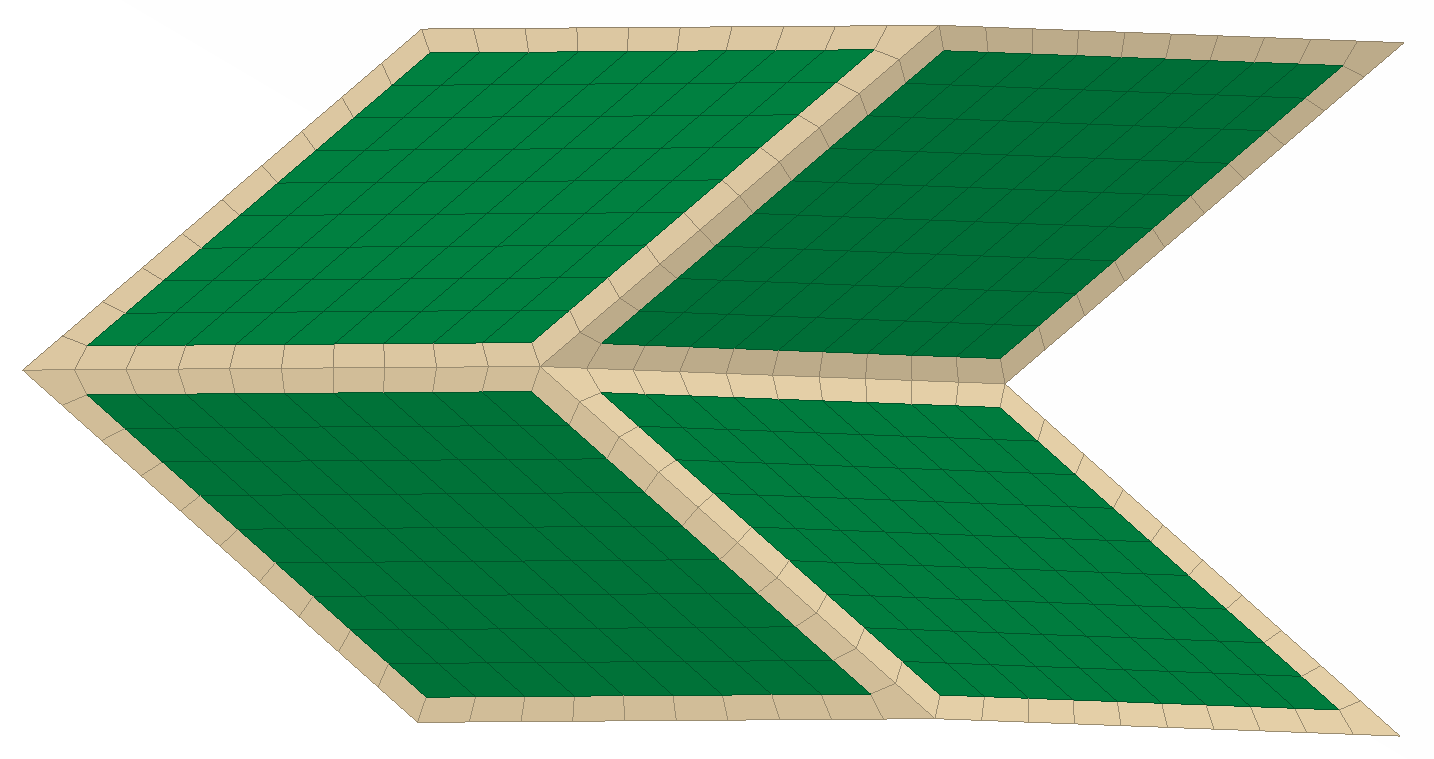}
	}
\caption{(a) Geometric parameters for a single unit cell of the Miura-Ori pattern  (b) Miura-Ori single unit cell without hinge (c) Miura-Ori single unit cell with hinge region (yellow color) }
\label{img:miura}
\end{figure}

Miura-Ori pattern consists of a 2D array of repeating units with four identical parallelograms in each unit, as shown in Figure~\ref{img:miura}(a). $l$ is the crease length, $\alpha$ is the angle between the sides of each parallelogram at the left-top corner, and $\beta$ is the angle between planes $S1$ and $S2$. $\theta$ is the angle between planes $S1$ and $S3$. Based on the paper by Wei et al.\cite{wei2013geometric}, the angles $\alpha$, $\beta$, and $\theta$ are correlated with each other and can be expressed with the Equation~\ref{eqn:ori_geom}.
\begin{equation}
\begin{aligned}
    \beta = 2 \,sin^{-1}[ \,\zeta \, sin(\theta/2)] \\
    l = 2 \,l_1 \,\zeta \\
    w = 2 \,l_2 \,\xi \\
    h = l_1\,\zeta \,tan\alpha \, cos(\theta/2) 
\end{aligned}
\label{eqn:ori_geom}
\end{equation}
\noindent where the dimensionless terms width and height can be expressed as: $\xi=sin\alpha \,sin(\theta/2)$ and $\zeta=cos\alpha \,(1-\xi^2)^{-1/2}$. Thus in this paper, only $\alpha$ and $\beta$ as considered as independent geometric parameters. We consider $\alpha$ values can vary between \ang{0} and \ang{90}, while $\beta$ values can vary between \ang{0} and \ang{180}. 

\begin{figure}[h!]
\centering
\subfigure[]{
	\includegraphics[width=0.5\textwidth]{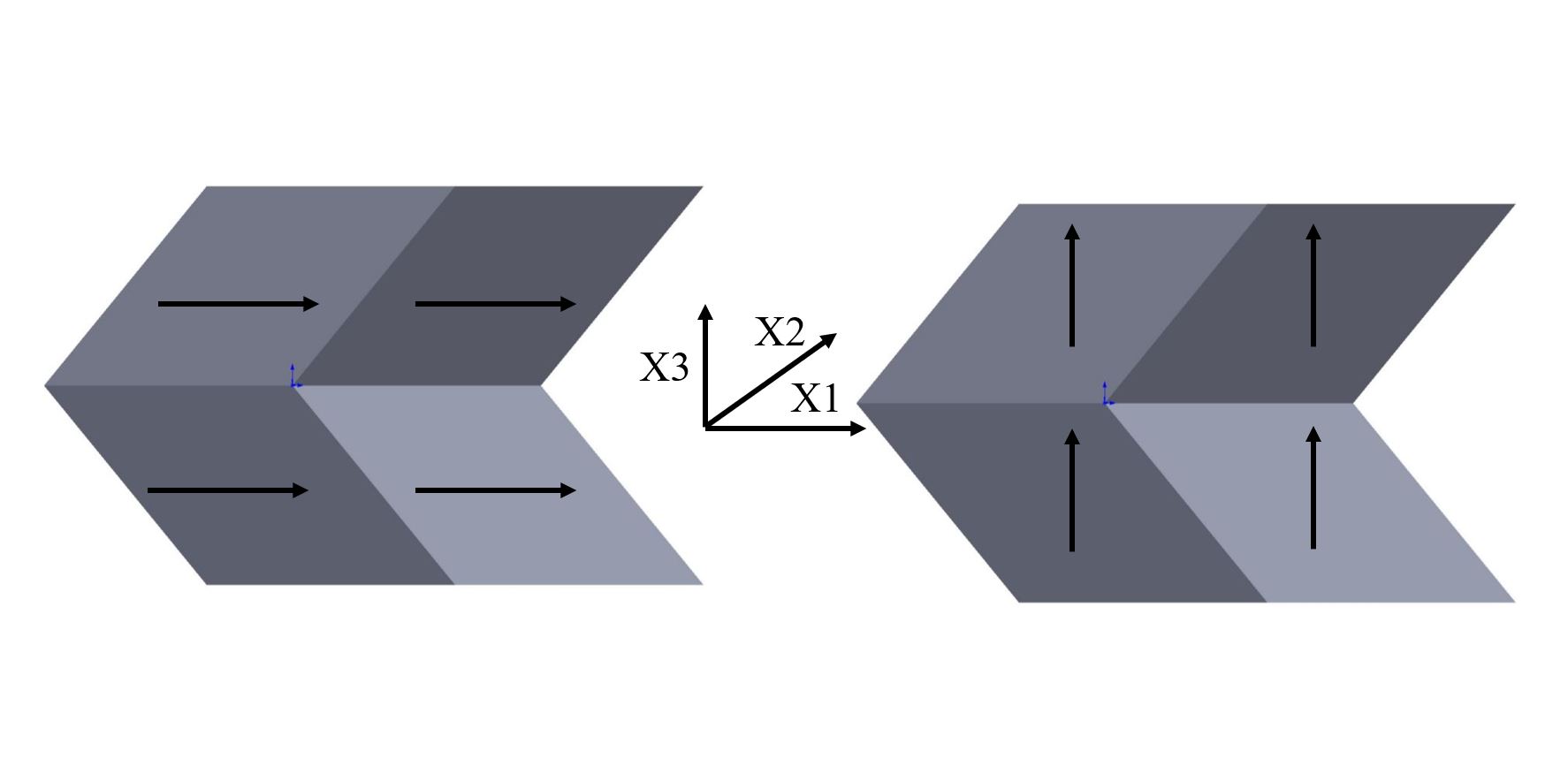}
	}
\centering
\subfigure[]{
	\includegraphics[width=0.35\textwidth]{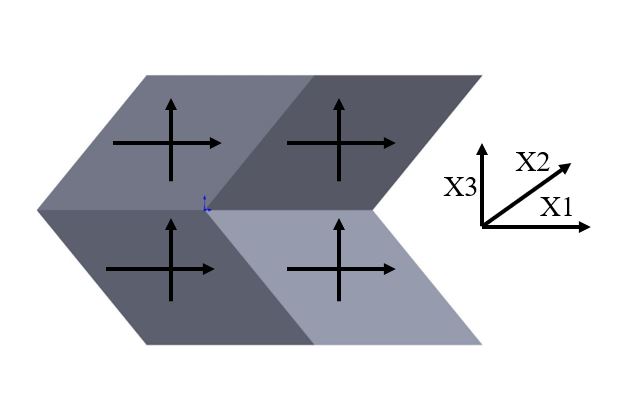}
	}	
	\caption{(a) Miura-Ori pattern with fiber direction along X1 axis - Case 1 and along X3 axis - Case 2 (b) Miura-Ori pattern with woven fiber reinforcements (fibers interlaced in two orthogonal directions - X1 and X3}
	\label{img:miura_direction}
\end{figure}

For directional Miura-Ori structures, three different fiber arrangements within each Miura-Ori unit are considered, as shown in Figure~\ref{img:miura_direction}. Fibers are either arranged along the $X1$ axis or $X3$ axis in all four parallelograms of a single unit (Figure~\ref{img:miura_direction}(a)), which can be achieved by using unidirectional fiber reinforced composites. The third case is where fibers are arranged in both $X1$ and $X3$ axes within each parallelogram (Figure~\ref{img:miura_direction}(b)), which can be achieved by a bi-directional woven fiber composite. Woven fiber composites are formed by weaving or interlacing warp and weft fiber bundles. Possible weaving architectures and choices of materials for warp/weft fiber bundles present a large range of possibilities that could influence in-plane mechanical responses, primarily in in-plane normal and shear directions. Due to the transversely isotropic behavior of fiber-reinforced composites in addition to the geometry of Miura-Ori units, the global mechanical properties of these structures are different in different directions. To study how these mechanical responses can be tuned, we study the compressive deformations in two principal directions - $X1$ and $X3$.

Miura-Ori patterns we have considered in this paper have dimensions of 10 mm x 10 mm for every parallelogram. These dimensions are based on that considered in Liu et al.\mbox{\cite{liu2015deformation}}, although these dimensions can be scaled. Each pattern consists of two regions: the cell region and the hinge region. We define two different types of arrangements: (1) both cell region and hinge region are made of the same stiff material like CFRP composites (we call it \textbf{rigid Miura-Ori}) - Figure~\ref{img:miura}(b); (2) cell region is made of a stiff material while the hinge region is made of soft material like silicone (we call it \textbf{flexible Miura-Ori}) - Figure~\ref{img:miura}(c). We compare the global mechanical responses of these two different types of Miura-Ori structures and explore how geometric and material parameters influence their mechanical responses. In this study, we set the width of the hinge region in the flexible Miura-Ori models to be 0.5 mm. For modeling purposes, we assume a perfectly joined connection between the hinge regions and carbon fiber-reinforced laminates in our geometry. From a manufacturing point of view, we can introduce flexible hinges by different methods. These methods are: (1) For a structure with continuous hinges, we can use a dual matrix hand layup approach to arrange the laminate parallelograms with fiber directions along specified directions and locations based on the design. This can be done by placing a stencil on a dry fabric’s hinge area and covering it with flexible resin, removing the stencil, and adding epoxy matrix everywhere else. (2) For a structure with discontinuous hinges, we can use mechanical hinges bonded/fastened to the stiff laminates. For rigid hinges, we can perform compression molding of polymer-infused carbon-reinforced fabric between molds of the desired design. 

\subsection{Material Properties}

Transversely isotropic linear elastic properties of unidirectional CFRP composites used in the Miura-Ori models shown in Figure~\ref{img:miura_direction}(a) are given in Table~\ref{tab:mat_prop}\cite{herakovich1998mechanics,PRABHAKAR2013-1}. For those with bi-directional materials shown in Figure~\ref{img:miura_direction}(b), we use the properties of woven fiber reinforced composites given in Table~\ref{tab:mat_prop}\cite{castellanos2018interlaminar}. In both cases, we consider homogenized laminate in this paper with effective properties given in \mbox{Table~\ref{tab:mat_prop}}. We do not explicitly model individual layers or fibers in this paper. To compare the response of our target Miura-Ori models which are made of CFRP or woven fiber composites, we consider a baseline model entirely made of an isotropic material - Aluminum. The Young's modulus and Poisson's ratio of Aluminum are $E=70GPa$ and $\mu=0.33$. For flexible Miura-Ori models with soft hinges, we use Young's modulus $E=3.1$ GPa and Poisson's ratio $\mu=0.48$\cite{resin}. An example of a single Miura-Ori unit with silicone resin in the hinge region is shown in Figure~\ref{img:miura}(c).

\begin{table}[h!]
\caption{Input material properties of Carbon Fiber Reinforced Polymer (CFRP) and Woven Composites}
\resizebox{\textwidth}{!}{
\begin{tabular}{cccccccccc}
\hline
Material Property & $E_{1}$      & $E_{2}$       & $E_{3}$       & $\mu_{12}$    & $\mu_{23}$    & $\mu_{13}$    & $G_{12}$      & $G_{23}$      & $G_{13}$      \\ \hline
CFRP Composite            & 155 GPa & 12.1 GPa & 12.1 GPa & 0.248 & 0.458 & 0.248 & 4.4 GPa & 3.2 GPa & 4.4 GPa \\
Woven Composite           & 85 GPa & 85 GPa & 12.1 GPa & 0.3 & 0.3 & 0.3 & 5 GPa & 0.765 GPa & 0.765 GPa \\ \hline
\end{tabular}
}
\label{tab:mat_prop}
\end{table}

\section{Computational Analysis} \label{sec:3_1}

This section describes the computational modeling of Miura-Ori structures under compression loading. We evaluate their response under external in-plane loading in the two in-plane principal directions.

\subsection{Model Parameters and Boundary Conditions}\label{sec:3_1}

\begin{figure}[h!]
\centering
\subfigure[]{
	\includegraphics[width=0.45\textwidth]{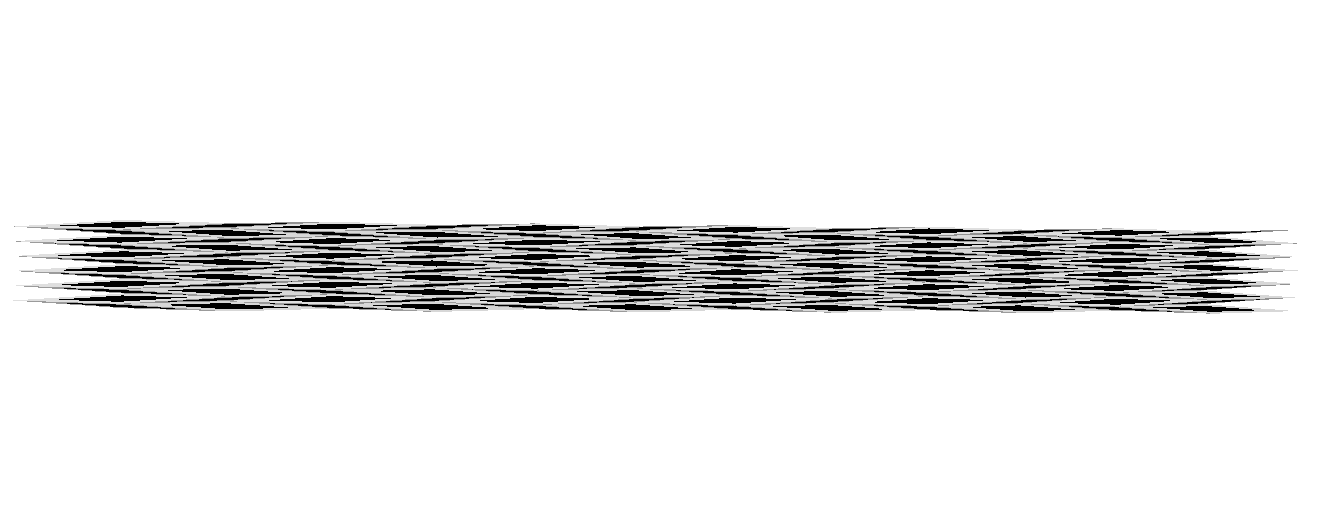}
	}
\centering
\subfigure[]{
	\includegraphics[width=0.45\textwidth]{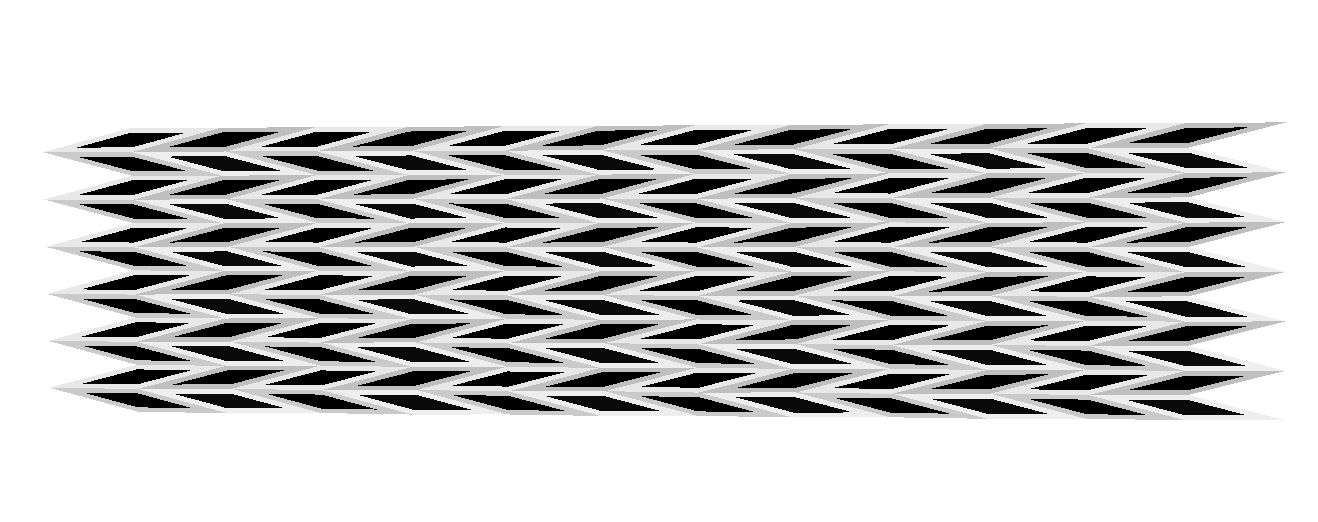}
	}
\centering
\subfigure[]{
	\includegraphics[width=0.45\textwidth]{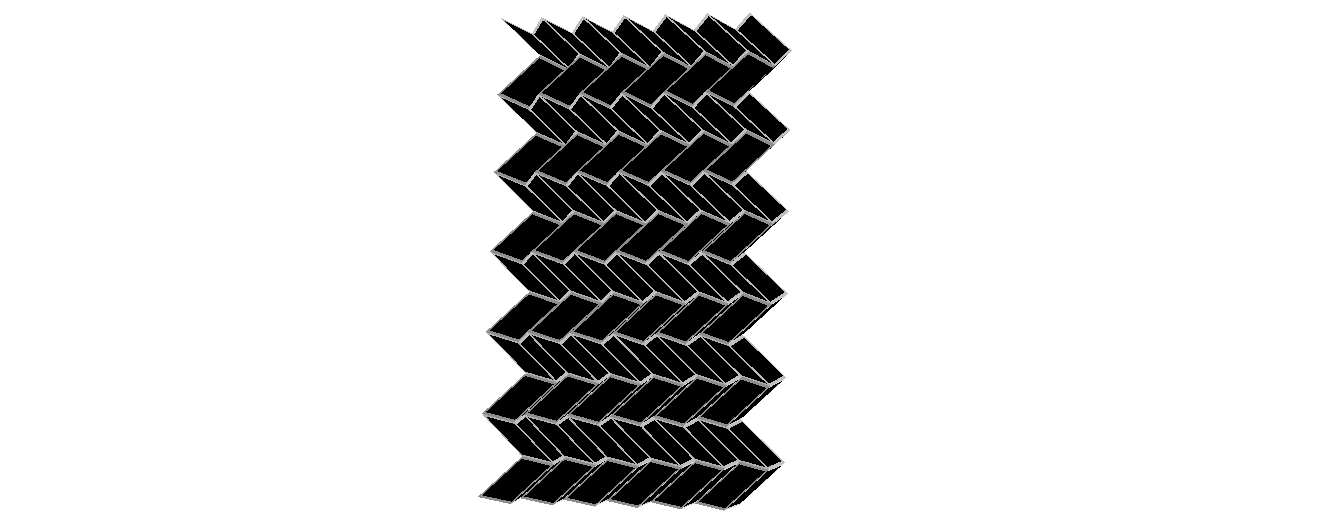}
	}	
\centering
\subfigure[]{
	\includegraphics[width=0.45\textwidth]{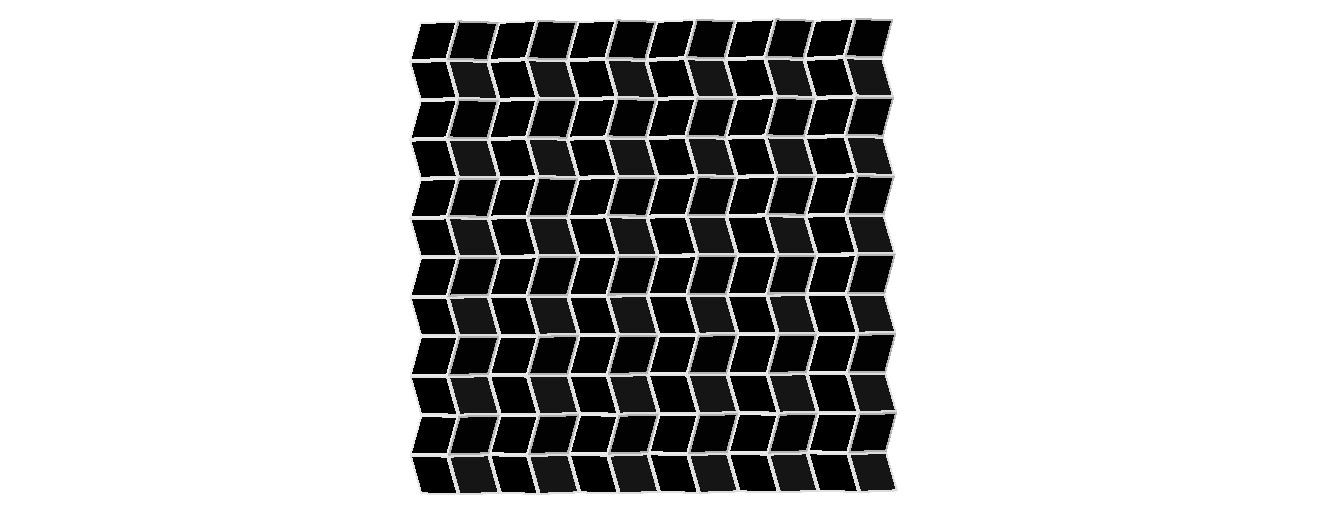}
	}	
\caption{Miura-Ori patterns with extreme values of $\alpha$ and $\beta$: (a) $\alpha=\ang{15}, \beta=\ang{30}$ (b) $\alpha=\ang{15}, \beta=\ang{150}$ (c) $\alpha=\ang{75}, \beta=\ang{30}$ (d) $\alpha=\ang{75}, \beta=\ang{150}$ (These figures showcase the shape of Miura-Ori patterns, and do not reflect the actual size)}
\label{img:geo_extreme}
\end{figure}

For compression analysis, we consider compressive load acting along each of the two orthogonal in-plane directions: X1 and X3. We determine two global mechanical responses of the Miura-Ori models: Poisson’s ratio and in-plane stiffness. We select combinations of $\alpha$ = (\ang{15}\ \ang{30}\ \ang{45}\ \ang{60}\ \ang{75}) and $\beta$ = (\ang{30}\ \ang{60}\ \ang{90}\ \ang{120}\ \ang{150}), and perform 25 simulations for each loading case. Miura-Ori patterns with combinations of extreme angles for $\alpha$ and $\beta$ - (\ang{15},\ang{30}), (\ang{15},\ang{150}), (\ang{75},\ang{30}), and (\ang{75},\ang{150}), are shown in Figure~\ref{img:geo_extreme}.

\begin{figure}[h!]
\centering
\includegraphics[width=0.5\textwidth]{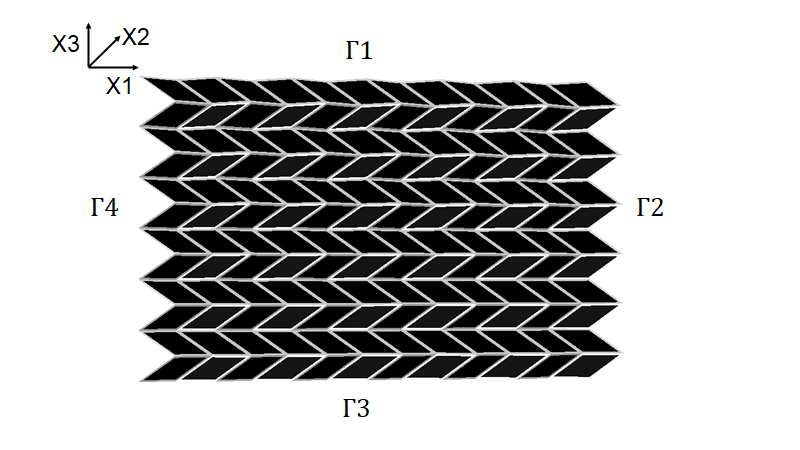}
\caption{Boundary definitions on Miura-Ori model for in-plane compression}

\label{img:miura_bc}
\end{figure}

Figure~\ref{img:miura_bc}(a) shows a Miura-Ori pattern with external boundaries defined as $\Gamma_1$, $\Gamma_2$, $\Gamma_3$, and $\Gamma_4$ for the top, right, bottom, and left edges. The boundary condition on each boundary is described below in terms of $u$ (X1) and $v$ (X3) displacements. Boundary conditions for compressive loading cases along X1 and X3 directions are described in Equation~\ref{eqn:bc_ver} and Equation~\ref{eqn:bc_hon}, respectively. Here, $d$ is the externally applied displacement set to 20\% of the domain length along the loading direction. This is to achieve enough transverse displacement to calculate the negative Poisson's ratio, although other displacement percentages can be chosen.

{
\vspace{0.1in}
\begin{equation}
\begin{split}
    v = \mathrm{constant} \quad \mathrm{on} \quad \Gamma_1 \\
    u = -d \quad \mathrm{on} \quad \Gamma_2 \\
    v = 0 \quad \mathrm{on} \quad \Gamma_3 \\
    u = 0 \quad \mathrm{on} \quad \Gamma_4
    \end{split}
\label{eqn:bc_ver}
\end{equation}
}

{
\vspace{0.1in}
\begin{equation}
\begin{split}
    v = -d \quad \mathrm{on} \quad \Gamma_1 \\
    u = \mathrm{constant} \quad \mathrm{on} \quad \Gamma_2 \\
    v = 0 \quad \mathrm{on} \quad \Gamma_3 \\
    u = 0 \quad \mathrm{on} \quad \Gamma_4
    \end{split}
\label{eqn:bc_hon}
\end{equation}
}

\subsection{Finite Element Modeling and Analysis}
Each model is subjected to a compressive displacement along either the vertical or the horizontal direction. Compressive stiffness is represented by a Compressive Stiffness Indicator (CSI) defined in Equation~\ref{eqn:stiff_ind} as the initial slope of the force-displacement graph.

\begin{equation}
    \mathrm{Compressive\: Stiffness\: Indicator,\: K_C} = \frac{\mathrm{Reaction\:Force\:at\:the\:Loading\:Edge}}{\mathrm{Loading\:Edge\:Displacement}}
\label{eqn:stiff_ind}
\end{equation}

Poisson's ratio is calculated using Equation~\ref{eqn:poisson}.

\begin{equation}
    \nu = -\frac{\epsilon_{\mathrm{transverse}}}{\epsilon_{\mathrm{axial}}}
\label{eqn:poisson}
\end{equation}

\noindent where, $\epsilon_{\mathrm{transverse}}=\frac{\Delta L_{\mathrm{transverse}}}{L_{\mathrm{transverse}}}$ and $\epsilon_{\mathrm{axial}}=\frac{\Delta L_{\mathrm{axial}}}{L_{\mathrm{axial}}}$. $L_{\mathrm{axial}}$ and $L_{\mathrm{transverse}}$ refer to the original length along the loading and the in-plane transverse directions. Correspondingly, $\Delta L_{axial}$ and $\Delta L_{transverse}$ are the length changes along these directions.

To determine the physical properties described above, a Miura-Ori model is first discretized by meshing with 3-node (S3) and 4-node (S4R) shell elements in a finite element software - ABAQUS\cite{0b112d0e5eba4b7f9768cfe1d818872e}. We perform a mesh convergence analysis to determine the appropriate mesh size. The connections between the hinge and CFRP regions are modeled as a rigid connection in our finite element model. After assigning the corresponding material properties, boundary conditions, and loads, this meshed structure is passed into a numerical analysis solver to determine the nodal displacements and related strain and stress fields.  

\section{Regression Analysis for Optimizing Miura-Ori Structures}\label{sec:regression}
To determine optimal geometry that can possess both high stiffness and negative Poisson's ratio, we perform multiple regression analysis to establish a relationship between geometric parameters ($\alpha$, $\beta$) and target effective properties. Multiple regression is a statistical method to determine the relationship between a single dependent variable and several independent variables. A regression coefficient $\beta$ is calculated by minimizing the sum of squared errors $\sum_{i} (Y_i-f(X_i,\beta))^2$, given the estimation function $f$ and $\{X_i, Y_i\}$ for each i-th sample from a dataset. In this paper, the independent variables are the geometric parameters ($\alpha$ and $\beta$), and the dependent variable is an effective target property (stiffness $K$ or Poisson's ratio $\nu$). Wei et al.\cite{wei2013geometric} established a relationship between effective stiffness $K$ and Poisson's ratio $\nu$ as a function of geometric angles in the trigonometric form. They utilized the Miura-Ori structure as our analysis but assumed a uniform material, the same edge length, and spring constant to derive these simplified analytical functions. However, in our analysis, as we consider the entire geometry with complex boundary conditions and directional materials, it is much more challenging to express $K$ and $\nu$ in simplified analytical functions. However, $K$ and $\nu$ are still functions related to $\alpha$ and $\beta$. As trigonometric functions can be represented with Taylor's series expansion, 
so we have the expression: $(K,\nu) \sim C + F_{0.5}(\alpha^{0.5},\beta^{0.5})+F_{1}(\alpha,\beta) + F_{1.5}(\alpha^{1.5},\beta^{1.5}) + F_{2}(\alpha^2,\beta^2) + ... + O(\alpha^n,\beta^n)$. Here, $F_{i}$ represents the function consisting of the i-th power of geometric parameters including cross-terms, and $C$ is the intercept term. Thus, we can use a linear regression model consisting of different orders of $\alpha$ and $\beta$ to approximate $K$ and $\nu$.

\subsection{Grid map interpolation}
\begin{figure}[h!]
\centering
	\includegraphics[width=0.5\textwidth]{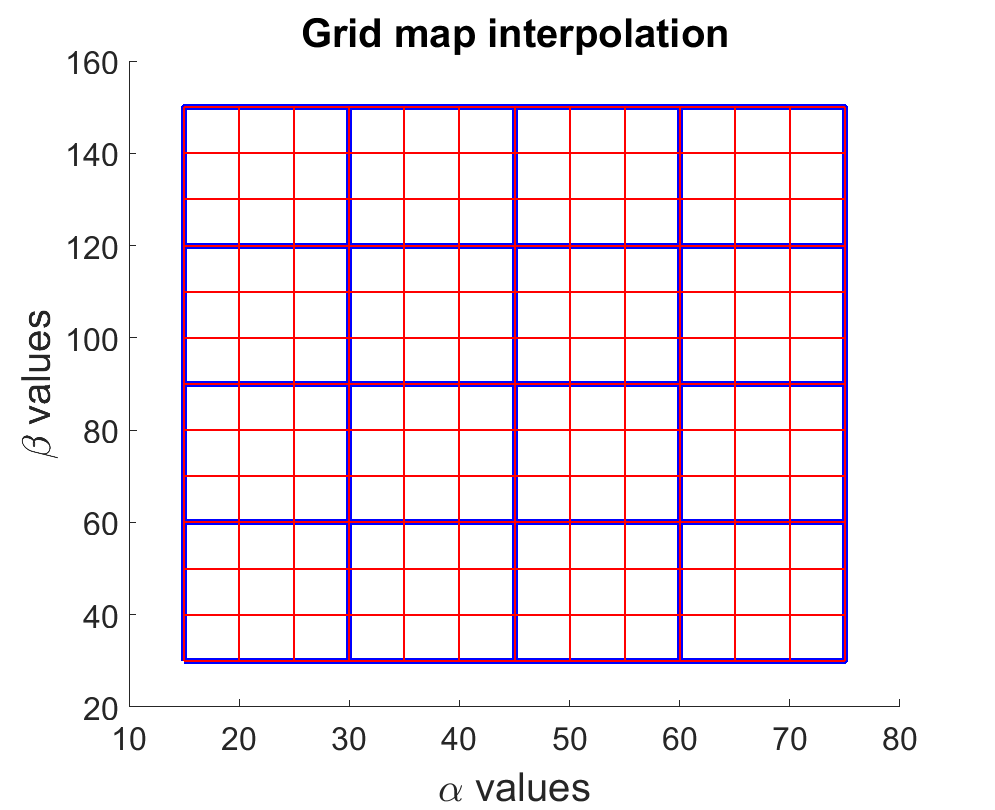}
\caption{Grid map differences: the blue margin is a 5-by-5 grid map formed by simulation results from FEA, and the red margin is the target 13-by-13 grid map}
\label{img:grid_map}
\end{figure}

In order to fit a regression model with higher-order terms, we first expand our data from FEA results using interpolation. We interpolate the original 5-by-5 contour (as discussed in Section~\ref{sec:3_1}) onto a map with a denser 13-by-13 grid for additional intermediate values of $\alpha$ and $\beta$. Figure~\ref{img:grid_map} shows the original 5-by-5 FEA grid in blue and the target 13-by-13 denser grid in red. In order to interpolate the original FEA data from the blue to the red grid, we utilize bilinear interpolation\cite{li2001new}. Within bilinear interpolation, the interpolated value for every red grid point is determined by using the blue grid point coordinates and values within the blue square that the red point lies in. The coordinates of the four vertices on a blue grid are denoted as $Q_{11}=(x_1,y_1)$, $Q_{12}=(x_1,y_2)$, $Q_{21}=(x_2,y_1)$, and $Q_{22}=(x_2,y_2)$, and corresponding values are denoted as $f(Q_{11})$, $f(Q_{12})$, $f(Q_{21})$, and $f(Q_{22})$. This bilinear interpolation takes the form shown in Equation~\ref{eqn:interpolation}.
 
\begin{equation}
\begin{aligned}
    &f(x,y) \approx \frac{y_2-y}{y_2-y_1}f(x,y_1) + \frac{y-y_1}{y_2-y_1}f(x,y_2) \\
    &= \frac{y_2-y}{y_2-y_1}\left[\frac{x_2-x}{x_2-x_1}f(Q_{11}) + \frac{x-x_1}{x_2-x_1}f(Q_{21})\right] + \frac{y-y_1}{y_2-y_1}\left[\frac{x_2-x}{x_2-x_1}f(Q_{12}) + \frac{x-x_1}{x_2-x_1}f(Q_{22})\right] \\
\end{aligned}
\label{eqn:interpolation}
\end{equation}

\subsection{Setup of Regression analysis}\label{sec:4_2}
With bilinear interpolation, we obtain the stiffness and Poisson's ratio contours with a denser grid and provide more data points for fitting the regression model. Since $(K,\nu) \sim C + F_{0.5}(\alpha^{0.5},\beta^{0.5})+F_{1}(\alpha,\beta) + F_{1.5}(\alpha^{1.5},\beta^{1.5}) + F_{2}(\alpha^2,\beta^2) + ... + O(\alpha^n,\beta^n)$, we consider up to a third order polynomial in each variable $\alpha$ and $\beta$ in order to select a regression model with reasonable complexity. For cross terms, we consider up to a second-order polynomial. However, the regression model is still complex and overfits the model. Hence, we conduct Stepwise Regression\cite{efroymson1960multiple} analysis to extract significant features and use adjusted R-squared ($\geq 0.75$) and P-values ($\alpha \leq 0.05$) as our selection criteria to select significant terms.

\section{Results and Discussion}

We first compare the global mechanical responses of flexible and rigid Miura-Ori structures for isotropic materials in Section~\ref{sec:rigid_flex}. Then in Section~\ref{sec:geom_param_regress}, we proceed with further in-depth studies on rigid Miura-Ori structures with different values of geometric parameters - $\alpha$ and $\beta$ and materials - Aluminum and CFRP composite. This helps us understand how CFRP composite models perform compared to Aluminum models, and how the optimal geometry differs. We use regression analysis to explore the relationships between geometric parameters and global mechanical responses. {We further optimize the Miura-Ori structures by proposing parameter $R$ that linearly combines different targets (like stiffness and Poisson's ratio), as shown in Section~\ref{sec:ori_optimize}. Finally in Section~\ref{sec:woven}, we analyze woven composite models by changing the weaving patterns to establish how individual material properties control the global mechanical behavior.

\subsection{Comparison between rigid Miura-Ori and flexible Miura-Ori structures}\label{sec:rigid_flex}
In order to compare the mechanical responses between rigid and flexible Miura-Ori structures, we subject the two models to two different load conditions: horizontal (X1) and vertical (X3) in-plane compression. Specifically, we choose the origami model with $\alpha=\ang{45}$ and $\beta=\ang{90}$ as a case study and compare the response between origami structures made of Aluminum and CFRP composites.

Table~\ref{tab:mat1} shows the comparison between stiffness and Poisson's ratio between rigid and flexible Miura-Ori structures. We observe that adding flexible resin in the Miura-Ori hinges will significantly reduce the stiffness and the absolute value of negative Poisson's ratio (NPR) in specific loading directions. Thus for the remainder of this paper, we primarily consider origami structures without resin, that is, rigid Miura-Ori structures.

\begin{table}[h!]
\centering
\caption{Mechanical properties of Miura-Ori models with (flexible) and without (rigid) resin at the hinges}
\resizebox{0.9\textwidth}{!}{%
\begin{tabular}{cccccc}
\hline
Loading Direction                 &         & Aluminum  & Aluminum with Resin & CFRP & CFRP with Resin\\ \hline
\multirow{2}{*}{In-plane X1 Load} & Max K (N/(mm))   & 8.42  & 1.59 & 2.36 & 0.842        \\  
                                  & Max NPR & -0.512    & -0.887  & -0.351  & -0.707            \\ 
\multirow{2}{*}{In-plane X3 Load} & Max K (N/(mm))  & 12.99  & 1.48 & 6.70 & 1.26        \\  
                                  & Max NPR & -0.395    & -0.395  & -0.577  & -0.399            \\ \hline
\end{tabular}
}
\label{tab:mat1}
\end{table}

\subsection{Influence of geometric parameters on global mechanical responses} \label{sec:geom_param_regress}
Next, we explore how the optimal stiffness and Poisson's ratio change with respect to different geometric angles within rigid Miura-Ori structures with isotropic (Aluminum) and directional (CFRP composite) material. To explore such behavior, we consider the mechanical loading of Miura-Ori models in horizontal in-plane compression (X1) and vertical in-plane compression (X3).
We calculate weight normalized properties, effective stiffness, and Poisson's ratio. For different values of angles $\alpha$ and $\beta$, the weight normalized value for any property is defined as $S_{norm}=S/m$, where $S$ is the property, and $m$ is the weight of the geometry. We then investigate optimal origami structures by comparing the normalized mechanical properties between isotropic and directional materials.

\subsubsection{Weight normalized stiffness and Poisson's ratio in compression along X1 axis}

Figure~\ref{img:aniso_com2_stiff} and Figure~\ref{img:aniso_com2_poi} show the weight normalized stiffness and Poisson's ratio maps for directional and isotropic Miura-Ori structures when compressed in the X1 direction. Specifically, to represent the negative Poisson's ratio in the Log scale, we calculate the Log scale of the absolute value of the negative Poisson's ratio and then add the negative sign. The horizontal and vertical axes in these figures are the geometric angles - $\alpha$ and $\beta$ that control model's shape. From Figure~\ref{img:aniso_com2_stiff} and Figure~\ref{img:aniso_com2_poi}, we observe the following:
\begin{enumerate}
\setlength{\itemsep}{0pt}
\item 
Figure~\ref{img:aniso_com2_stiff}(a) shows the weight normalized stiffness contour for the isotropic case, where we observe that the optimal geometry corresponding to maximum stiffness occurs when $\alpha$ approaches \ang{0} and $\beta$ approaches \ang{180}. As $\alpha$ decreases, the structure is compact in the X3 direction, which is perpendicular to X1. While, as $\beta$ increases, the structure approaches a flat surface. Consequently, these two scenarios in combination result in a stiffer structure when compressed along the X1 direction. On the other hand, we notice that the optimal geometry that manifests the maximum value of negative weight normalized Poisson's ratio corresponds to $\alpha$ and $\beta$ approaching \ang{0}. 
\item Compared to the isotropic model, the directional model has a similar optimal shape with respect to both stiffness and Poisson's ratio. Moreover, when fiber direction is in line with the loading direction, as in \textbf{case 1} shown in Figure~\ref{img:miura_direction}(a), we observe that the directional model can manifest higher magnitudes of maximum normalized stiffness and negative Poisson's ratio compared to the isotropic model when compressed along X1 direction.
\end{enumerate}

\begin{figure}[h!]
\centering
\subfigure[]{
	\includegraphics[width=0.31\textwidth]{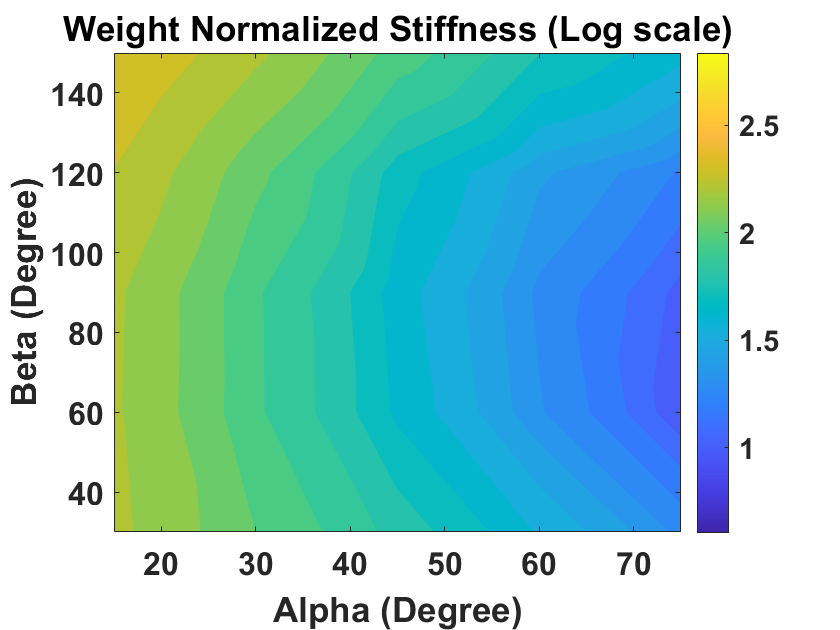}
	}
\centering
\subfigure[]{
	\includegraphics[width=0.31\textwidth]{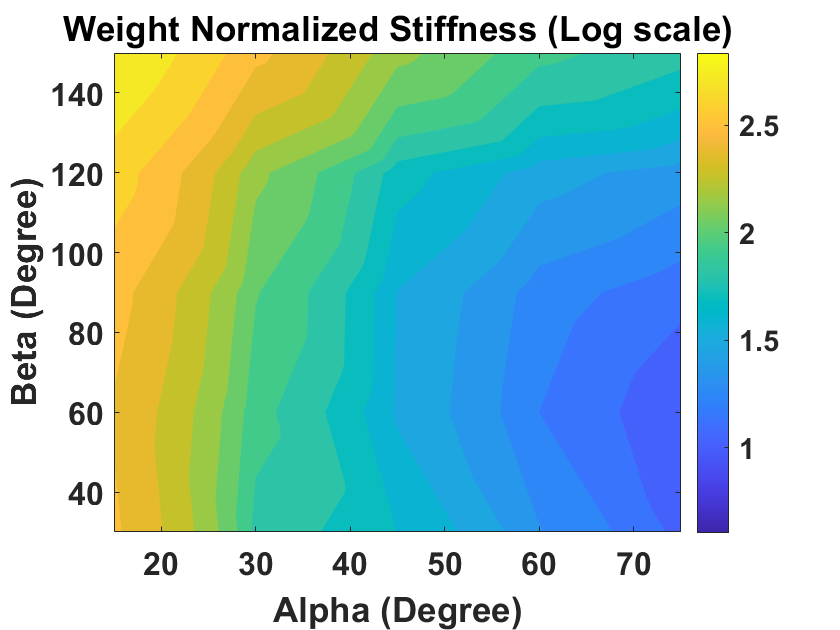}
	}
\centering
\subfigure[]{
	\includegraphics[width=0.31\textwidth]{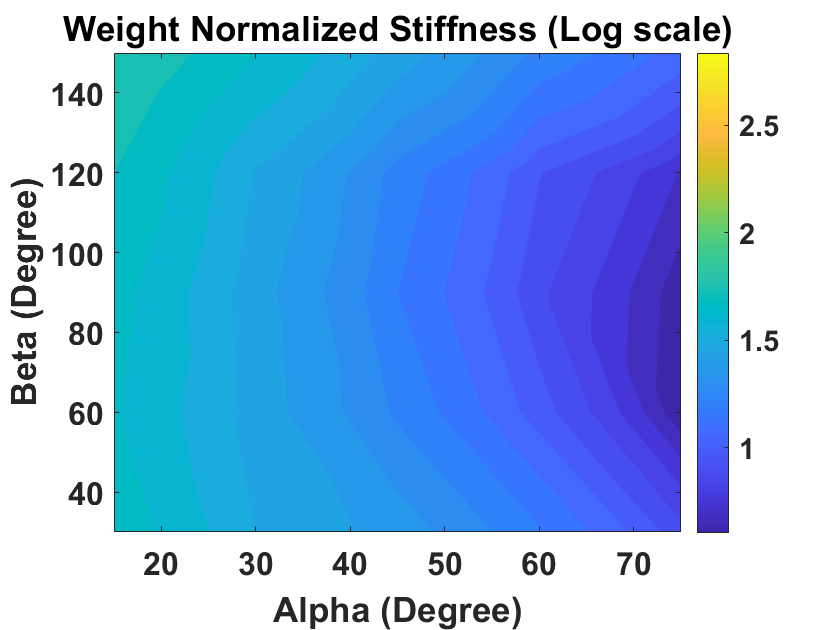}
	}	
	\caption{Weight normalized stiffness values ($N/(mm*g)$) of Miura-Ori models when compressed along X1: (a) isotropic case (b) fiber direction Case 1 (c) fiber direction Case 2}
	\label{img:aniso_com2_stiff}
\end{figure}

\begin{figure}[h!]
\centering
\subfigure[]{
	\includegraphics[width=0.31\textwidth]{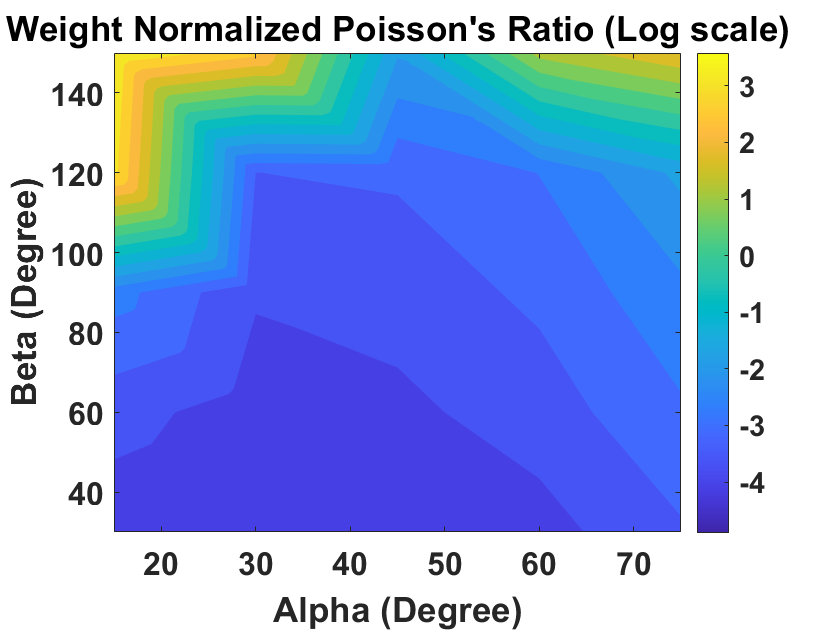}
	}
\centering
\subfigure[]{
	\includegraphics[width=0.31\textwidth]{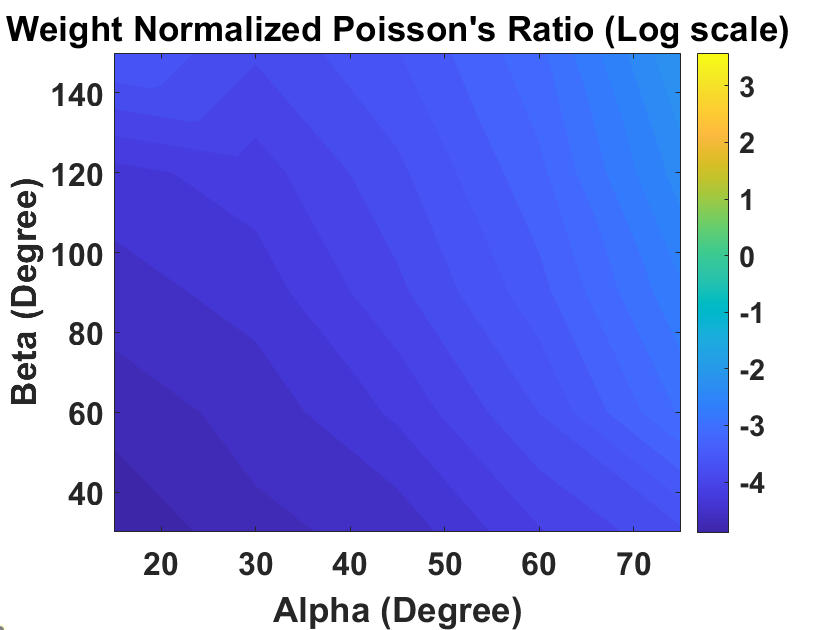}
	}
\centering
\subfigure[]{
	\includegraphics[width=0.31\textwidth]{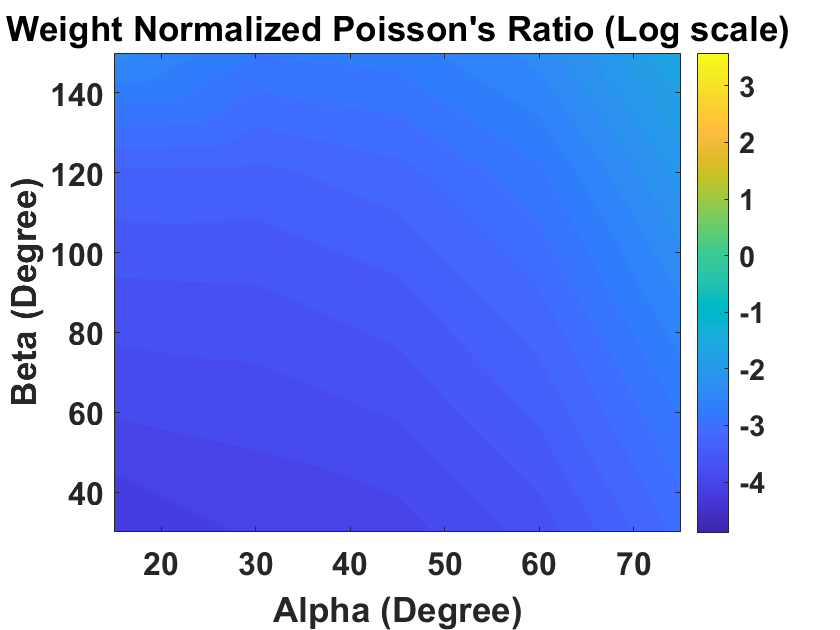}
	}	
	\caption{Weight normalized Poisson's ratio values ($1\times 10^3/g$) of Miura-Ori models when compressed along X1: (a) isotropic case (b) fiber direction case 1 (c) fiber direction case 2}
	\label{img:aniso_com2_poi}
\end{figure}

\subsubsection{Weight normalized stiffness and Poisson's ratio in compression along X3 axis} \label{sec:compress_X1}

Figure~\ref{img:aniso_com1_stiff} and Figure~\ref{img:aniso_com1_poi} show the weight normalized stiffness and Poisson's ratio values for directional and isotropic Miura-Ori structures when compressed along X3 direction. 

From Figures~\ref{img:aniso_com1_stiff} and ~\ref{img:aniso_com1_poi}, we observe the following:
\begin{enumerate}
\setlength{\itemsep}{0pt}
\item From the stiffness contour for the isotropic model shown in Figure~\ref{img:aniso_com1_stiff}(a), we observe that the optimal geometry that has the maximum stiffness corresponds to $\alpha$ approaching \ang{0} and $\beta$ approaching \ang{180}. This optimal geometry is similar to that determined for the previous loading direction: compression along X1. Figure~\ref{img:geo_extreme}(b) shows an example Miura-Ori structure with very small $\alpha$ (\ang{15}) and large $\beta$ (\ang{150}). 
As for Poisson's ratio, we notice that the optimal geometry to possess a maximum value of negative weight normalized Poisson's ratio corresponds to $\alpha$ approaching \ang{90} and $\beta$ approaching \ang{60} as shown in Figure~\ref{img:aniso_com1_poi}(a). 
\item Compared to the isotropic model, the directional model (fiber direction case 1 and case 2) has similar optimal geometric parameters for both stiffness and Poisson's ratio values. However, we notice that the material direction in Miura-Ori structures dramatically influences on the value of weight normalized stiffness and Poisson's ratio. When the fiber direction is in the same direction of load, as in \textbf{case 2} shown in Figure~\ref{img:miura_direction}(a), directional Miura-Ori structures can effectively increase both maximum weights normalized stiffness and negative Poisson's ratio in the X3 direction as compared to the isotropic Miura-Ori structure. 
\end{enumerate}

\begin{figure}[h!]
\centering
\subfigure[]{
	\includegraphics[width=0.31\textwidth]{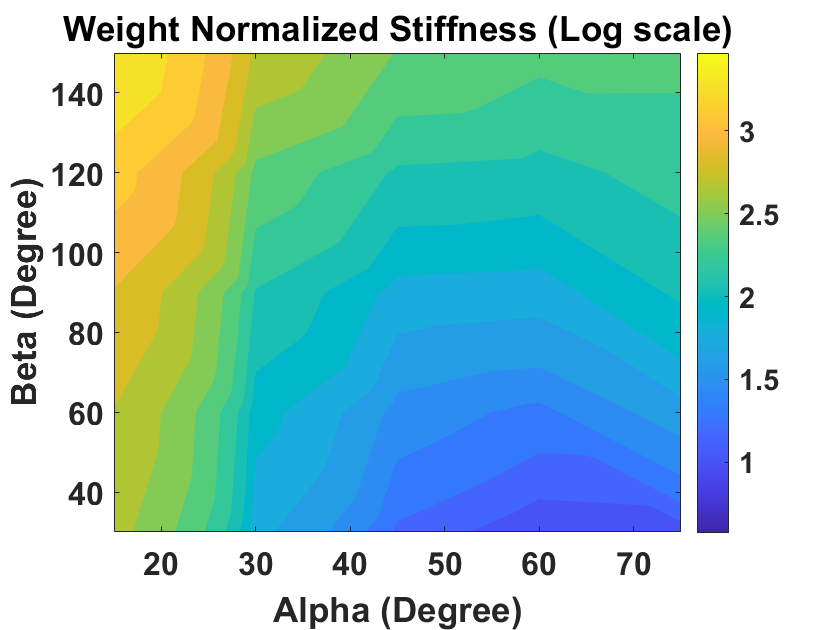}
	}
\centering
\subfigure[]{
	\includegraphics[width=0.31\textwidth]{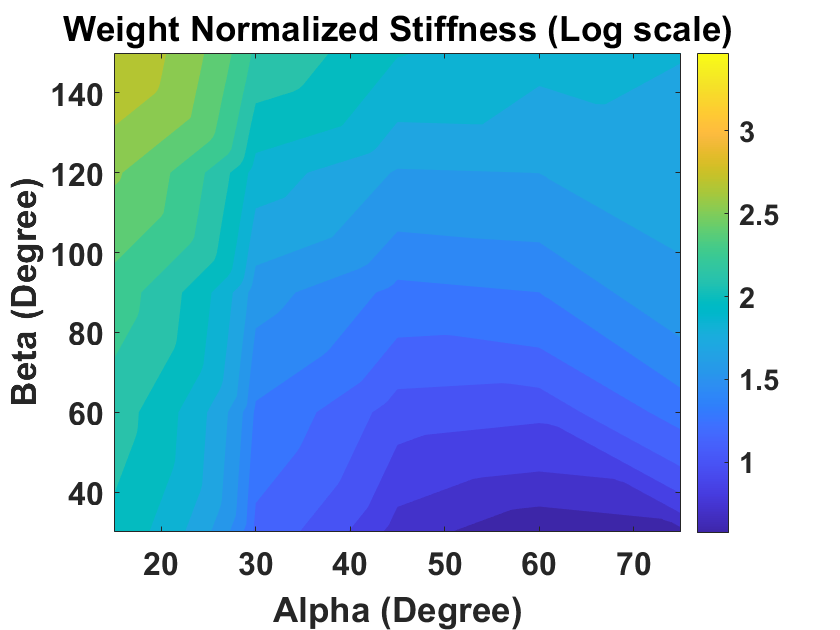}
	}
\centering
\subfigure[]{
	\includegraphics[width=0.31\textwidth]{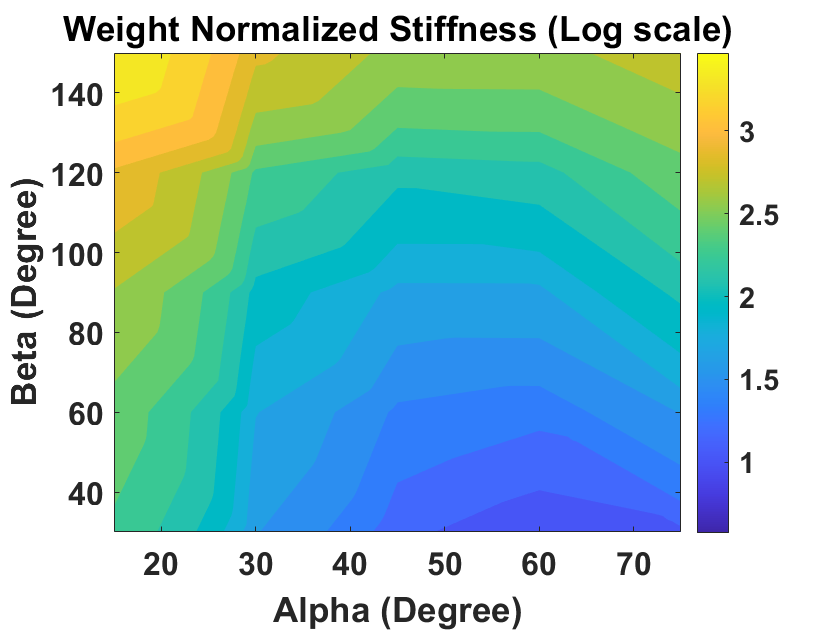}
	}	
	\caption{Weight normalized stiffness values ($N/(mm*g)$) of Miura-Ori models when compressed along X3: (a) isotropic case (b) fiber direction case 1 (c) fiber direction case 2}
	\label{img:aniso_com1_stiff}
\end{figure}

\begin{figure}[h!]
\centering
\subfigure[]{
	\includegraphics[width=0.31\textwidth]{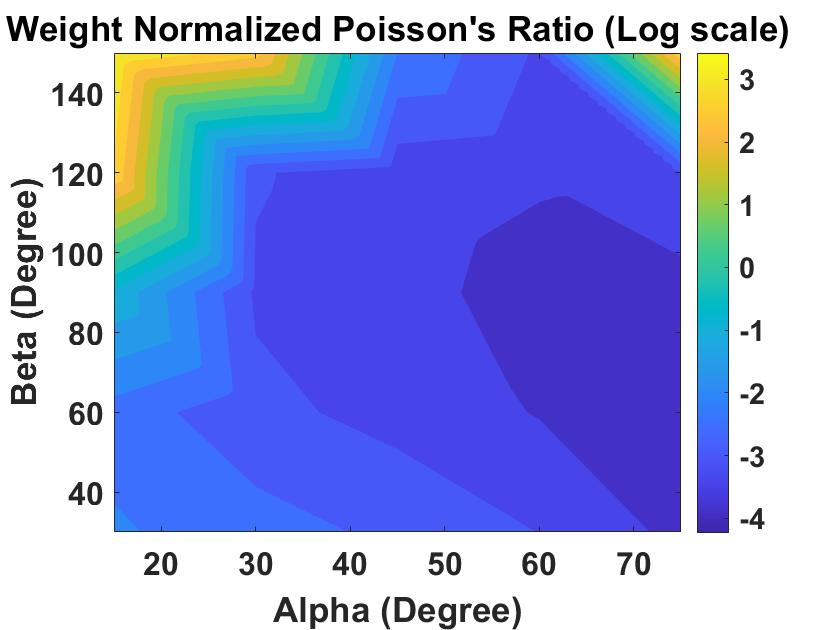}
	}
\centering
\subfigure[]{
	\includegraphics[width=0.31\textwidth]{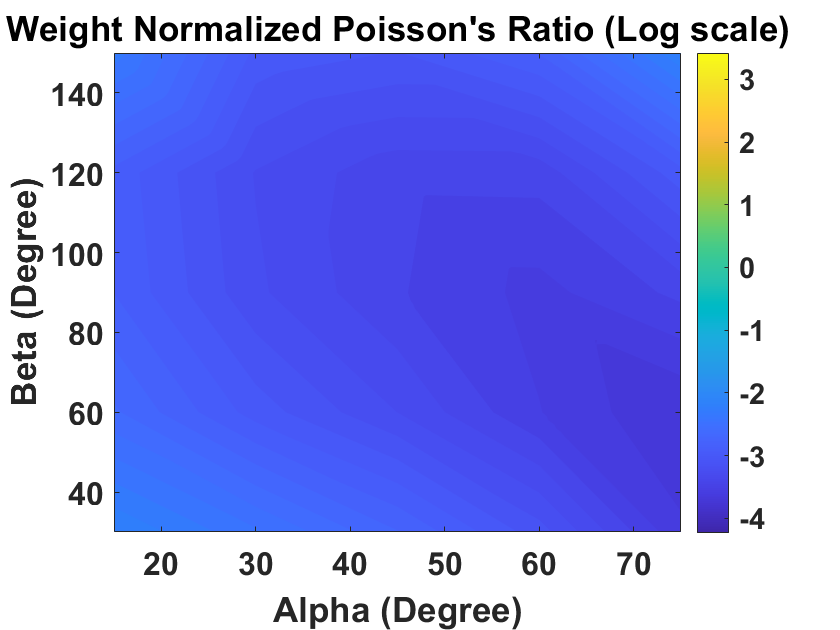}
	}
\centering
\subfigure[]{
	\includegraphics[width=0.31\textwidth]{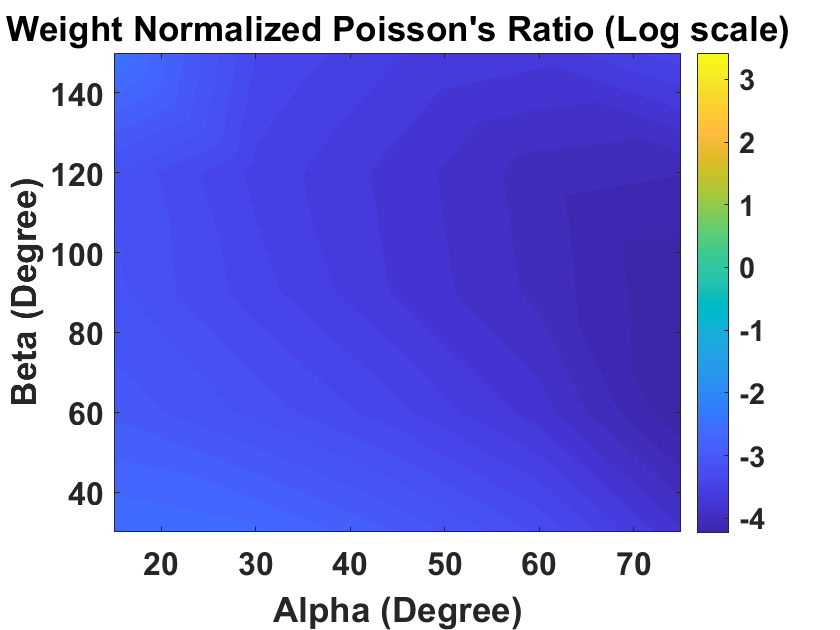}
	}	
	\caption{Weight normalized Poisson's ratio values ($1\times 10^3/g$) of Miura-Ori models when compressed along X3: (a) isotropic case (b) fiber direction case 1 (c) fiber direction case 2}
	\label{img:aniso_com1_poi}
\end{figure}

\subsection{Miura-Ori structure optimization based on regression analysis} \label{sec:ori_optimize}

In order to optimize the Miura-Ori structure, we introduce a regression analysis approach to mathematically represent the stiffness and Poisson's ratio of Miura-Ori structures as described in Section~\ref{sec:regression}. In this section, we will first validate the accuracy of our proposed regression model by comparing it to FEA results. We will then use our regression model to establish the optimal geometric parameters of the Miura-Ori structure based on different stiffness and Poisson's ratio values.

\subsubsection{Regression model selection and performance prediction}

From the Stepwise Regression analysis discussed in Section~\ref{sec:4_2}, we observe that nine parameters related to $\alpha$ and $\beta$ are significant for representing the regression model. However, after setting up the regression model, we noticed that parameters related to $\alpha^{0.5}$ could cause large over-fitting issues. Thus we keep seven significant parameters and set up our regression model as Equation~\ref{eqn:regression}.

\begin{equation}
\begin{split}
    K   = w_{01}+w_{11}\alpha+w_{21}\alpha^2+w_{31}\alpha^3+w_{41}\beta+w_{51}\beta^2+w_{61}\beta^3+w_{71}\alpha\beta\\
    \nu = w_{02}+w_{12}\alpha+w_{22}\alpha^2+w_{32}\alpha^3+w_{42}\beta+w_{52}\beta^2+w_{62}\beta^3+w_{72}\alpha\beta\\
\end{split}
\label{eqn:regression}
\end{equation}
where $w_{ij}$ refer to different regression coefficients.

Figure~\ref{img:regression_interp1} and \ref{img:regression_interp2} show examples of how the regression model predicts the contour compared to our FEA results on stiffness and Poisson's ratio when loading is applied in the X1 direction. Comparing the contours (a)-(b) and (c)-(d) in the two figures, we notice that our regression model could effectively capture the contour trend and the maximum/minimum locations and also provide a good prediction of the nodal values.

\begin{figure}[h!]
\centering
\subfigure[]{
	\includegraphics[width=0.41\textwidth]{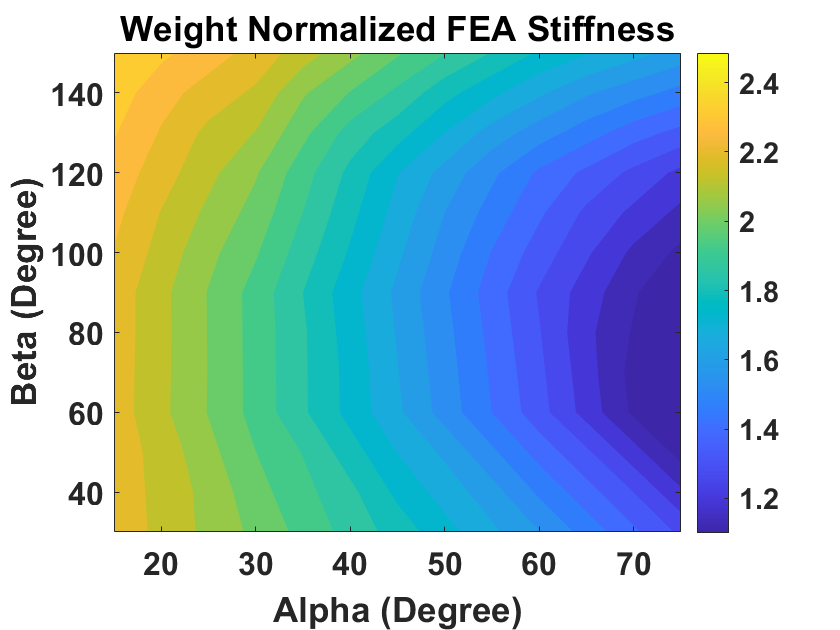}
	}
\centering
\subfigure[]{
	\includegraphics[width=0.41\textwidth]{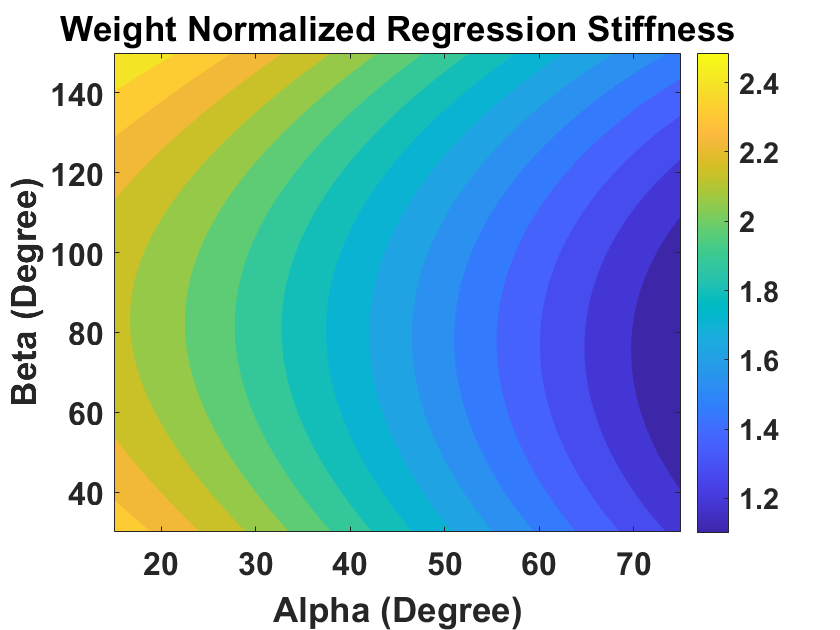}
	}
\centering
\subfigure[]{
	\includegraphics[width=0.41\textwidth]{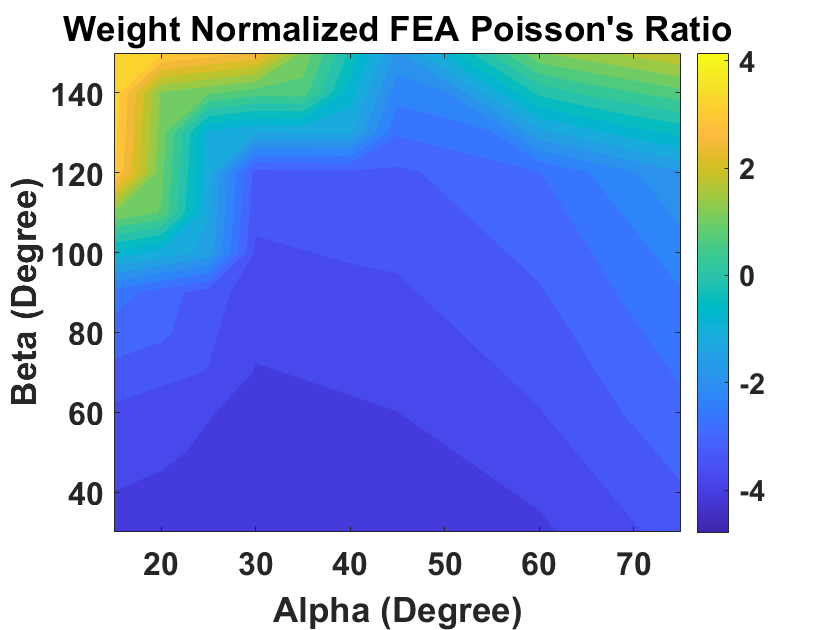}
	}
\centering
\subfigure[]{
	\includegraphics[width=0.41\textwidth]{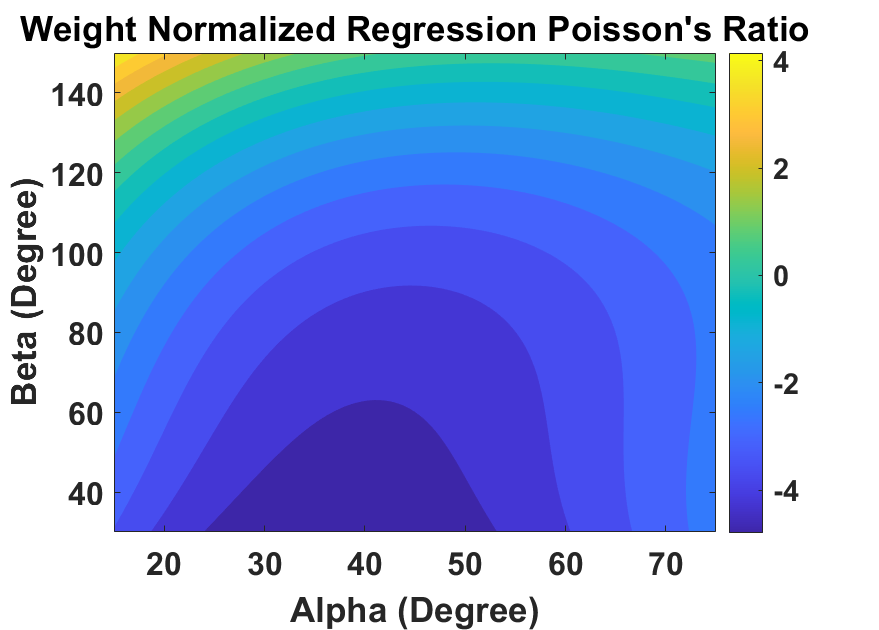}
	}	
	\caption{Comparison of regression model prediction and FEA results of Miura-Ori structures with isotropic material: (a) FEA contour and (b) regression model predicted contour for weight normalized stiffness ($N/(mm*g)$); (c) FEA contour and (d) regression model predicted contour for weight normalized Poisson's ratio ($1\times 10^3/g$). Compressive loading was applied in the X1 direction.}
	\label{img:regression_interp1}
\end{figure}

\begin{figure}[h!]
\centering
\subfigure[]{
	\includegraphics[width=0.41\textwidth]{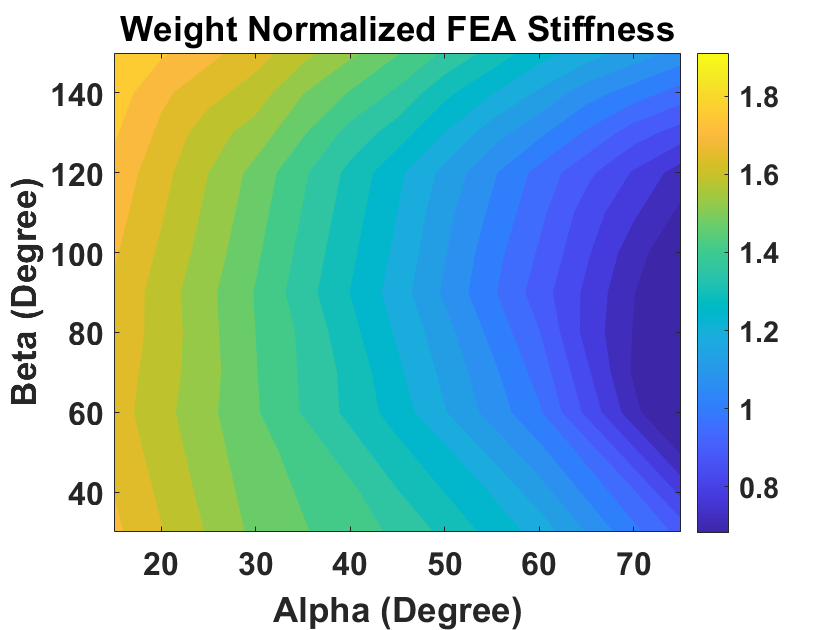}
	}
\centering
\subfigure[]{
	\includegraphics[width=0.41\textwidth]{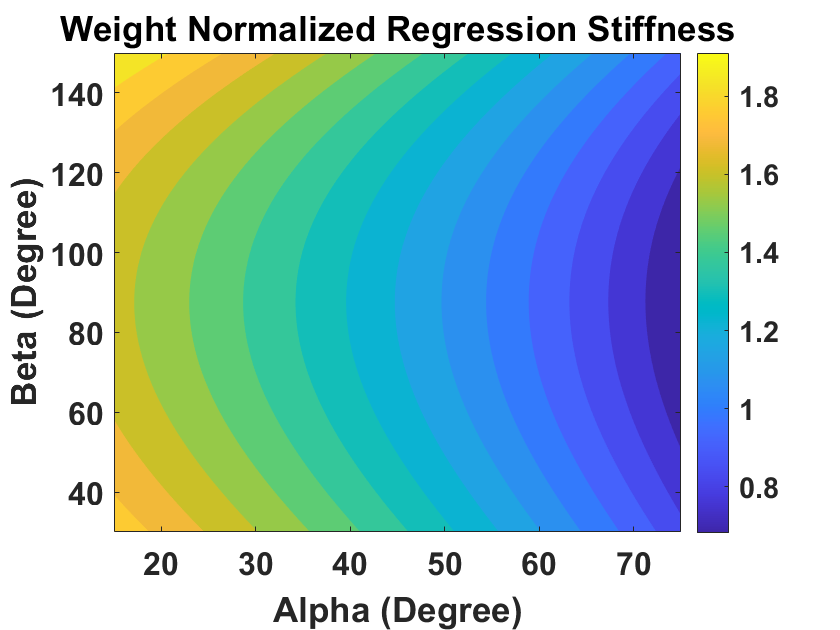}
	}
\centering
\subfigure[]{
	\includegraphics[width=0.41\textwidth]{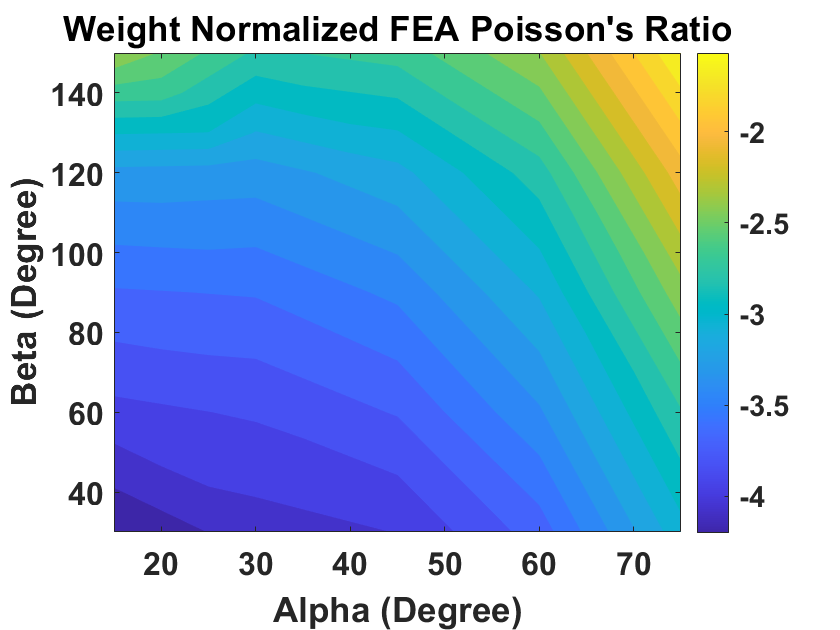}
	}
\centering
\subfigure[]{
	\includegraphics[width=0.41\textwidth]{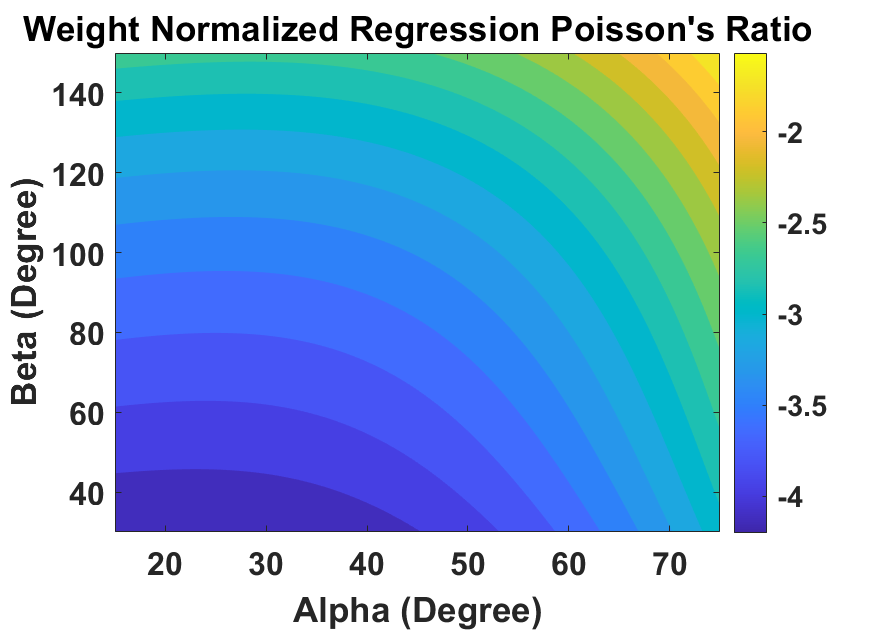}
	}	
	\caption{Comparison of regression model prediction and FEA results of Miura-Ori structures with directional material: (a) FEA contour and (b) regression model predicted contour for weight normalized stiffness ($N/(mm*g)$); (c) FEA contour and (d) regression model predicted contour for weight normalized Poisson's ratio ($1\times 10^3/g$). Compressive loading was applied in the X1 direction.}
	\label{img:regression_interp2}
\end{figure}

\subsubsection{Optimal Miura-Ori structure considering stiffness and Poisson's ratio}
After obtaining the regression model, we further utilize the model to determine optimal geometry with respect to the model's stiffness and Poisson's ratio. To set up a model combining different regression models for stiffness and Poisson's ratio, we define an abstract parameter R, which is the linear combination of two normalized parameters: $R=c_{1}\frac{K}{|K|_{max}}+(-1)^n(1-c_{1})\frac{\nu}{|\nu|_{max}}$, where $c_1\in [0,1]$, is a coefficient that reflects the weights between stiffness and Poisson's ratio and can be arbitrarily adjusted. For example, $c_1$ can be set to a smaller value if we focus on Poisson's ratio's optimization. $n$ is a value that controls the sign of the second term: $n=1$ when we want maximum stiffness and maximum negative Poisson's ratio, and $n=2$ when we want maximum stiffness and maximum positive Poisson's ratio. $|K|_{max}$ and $|\nu|_{max}$ are the max absolute values of the stiffness and Poisson's ratio contours, which are constants and can be set as a value based on FEA analysis or user's knowledge.

To determine the optimal geometry, we take the first derivative of the function $R$ and set it to zero, as shown in Equation~\ref{eqn:regression_derivative_R}. 

\begin{equation}
\begin{split}
    R_{\alpha} = \frac{\partial R}{\partial \alpha} = \frac{c_1}{|K|_{max}}\frac{\partial K}{\partial \alpha}+(-1)^n\frac{1-c_1}{|\nu|_{max}}\frac{\partial \nu}{\partial \alpha}=0 \\ 
    R_{\beta}  = \frac{\partial R}{\partial \beta} =\frac{c_1}{|K|_{max}}\frac{\partial K}{\partial \beta}+(-1)^n\frac{1-c_1}{|\nu|_{max}}\frac{\partial \nu}{\partial \beta}=0
\end{split}
\label{eqn:regression_derivative_R}
\end{equation}

Equation~\ref{eqn:regression_derivative_R} generates a system of equations with respect to $\alpha$ and $\beta$. Pairs of $\alpha$ and $\beta$ could uniquely determine the Miura-Ori pattern. For example, here, we consider the weights between stiffness and Poisson's ratio to be the same, that is, $c_1=0.5$. In Figure~\ref{img:R_contour}, we plot the contour of R for isotropic and directional Miura-Ori structures (where fiber and compression loading are in X3 direction) and calculate the analytical solution from the expression of R using Equation~\ref{eqn:regression_derivative_R}; it turns out the analytical solution matches with contour generated from discrete FEA data. This further proves that our proposed construction of abstract parameter $R$ can linearly combine optimization requirements on both stiffness and Poisson's ratio analytically and numerically.

\begin{figure}[h!]
\centering
\subfigure[]{
	\includegraphics[width=0.41\textwidth]{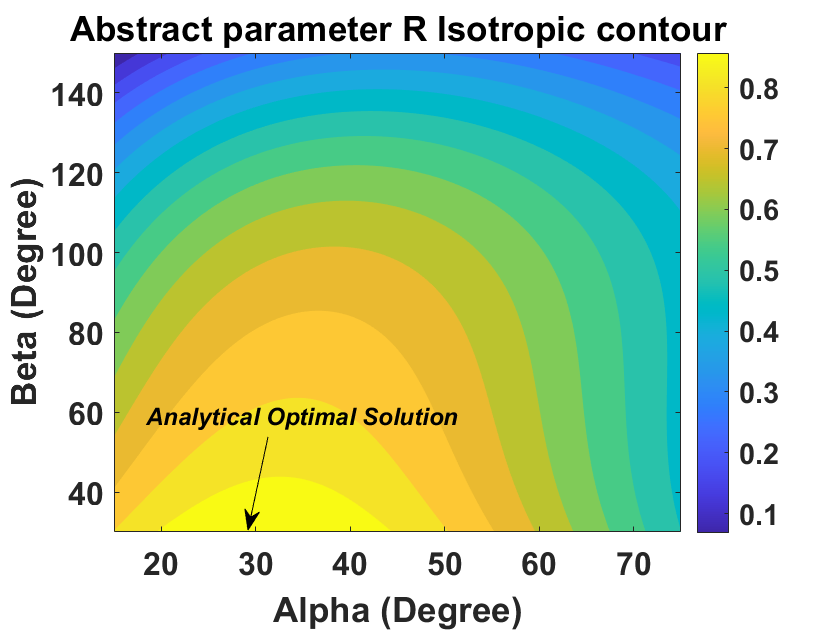}
	}
\centering
\subfigure[]{
	\includegraphics[width=0.41\textwidth]{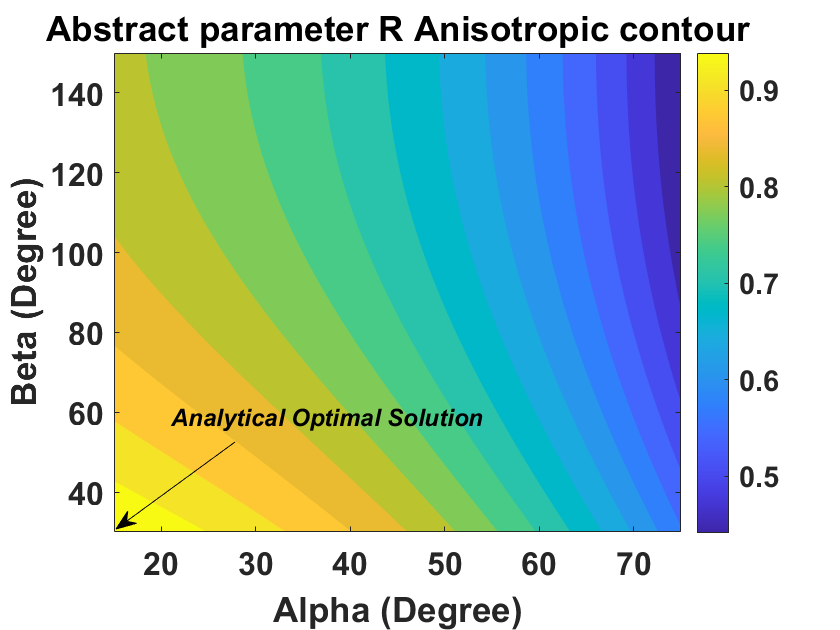}
	}
\caption{Map of the abstract parameter R for Miura-Ori structures under compression loading in the X3 direction with (a) an isotropic material model and (b) a directional material model (fiber in the X3 direction). Also shown is the optimal geometry determined using the contour and regression model analytical solution.}
\label{img:R_contour}
\end{figure}

\subsection{Influence of individual material property on Miura-Ori structures}\label{sec:woven}

To understand how different material properties affect the Miura-Ori structure given fixed geometric parameters $\alpha$ and $\beta$, we consider several structures with fixed geometries and test how different materials change the structure's mechanical responses. For example, we consider different combinations of $\alpha=\ang{30}$ $\beta=\ang{60}$, $\alpha=\ang{30}$ $\beta=\ang{90}$, $\alpha=\ang{45}$ $\beta=\ang{90}$, $\alpha=\ang{75}$ $\beta=\ang{60}$ and $\alpha=\ang{75}$ $\beta=\ang{90}$. 

Here we show an example of a Miura-Ori structure with $\alpha=\ang{45}$ and $\beta=\ang{90}$, but the same conclusion is valid for other combinations mentioned above. First, we investigate how a composite's individual material property controls the Miura-Ori structure's mechanical response. We consider six different artificial materials as shown in Table~\ref{tab:art_origami} where we change individual material properties of the CFRP composites. Here, we envision that the Miura-Ori structure could be made of woven composites since we can change its in-plane responses with different weave patterns and yarn materials. This enables us to investigate how in-plane material property influences the corresponding mechanical responses by generating a series of artificial materials. Hence, Miura-Ori structures are generated based on woven CFRP composite, and we substitute some material properties with Aluminum to consider artificial woven materials. Table~\ref{tab:mat2} shows the Miura-Ori structure's  mechanical responses for this geometry combination ($\alpha,\beta=\ang{45},=\ang{90}$) under X1 direction load for different artificial materials.

\begin{table}[h!]
\centering
\caption{Input material properties of artificial woven materials}
\resizebox{0.8\textwidth}{!}{%
\begin{tabular}{cccccccccc}
\hline
                                                                                    & $E_1$ (GPa) & $E_2$ (GPa) & $E_3$ (GPa)  & $v_{12}$  & $v_{23}$  & $v_{13}$  & $G_{12}$ (GPa)  & $G_{23}$ (GPa)   & $G_{13}$ (GPa)   \\ \hline
\begin{tabular}[c]{@{}c@{}}CFRP \\ Woven \end{tabular}  & 85  & 85  & 12.1 & 0.3  & 0.3  & 0.3  & 5    & 0.765 & 0.765 \\ 
\begin{tabular}[c]{@{}c@{}}Artificial \\ Woven 1\end{tabular}  & 70  & 70  & 70   & 0.3 & 0.3 & 0.3 & 5    & 0.765 & 0.765 \\ 
\begin{tabular}[c]{@{}c@{}}Artificial \\ Woven 2\end{tabular} & 70 & 70 & 70 & 0.33 & 0.33 & 0.33 & 5    & 0.765 & 0.765 \\ 
\begin{tabular}[c]{@{}c@{}}Artificial \\ Woven 3\end{tabular}  & 70 & 70 & 70 & 0.3  & 0.3  & 0.3  & 26.9 & 26.9  & 26.9  \\ 
\begin{tabular}[c]{@{}c@{}}Artificial \\ Woven 4\end{tabular} & 85  & 85  & 12.1 & 0.3  & 0.3  & 0.3  & 26.9 & 26.9  & 26.9  \\ 
\begin{tabular}[c]{@{}c@{}}Artificial \\ Woven 5\end{tabular} & 85  & 85  & 12.1 & 0.33  & 0.33  & 0.33  & 26.9 & 26.9  & 26.9  \\ 
\begin{tabular}[c]{@{}c@{}}Artificial \\ Woven 6\end{tabular} & 85  & 85  & 12.1 & 0.4  & 0.4  & 0.4  & 5 & 0.765  & 0.765 \\ \hline
\end{tabular}%
}
\label{tab:art_origami}
\end{table}

\begin{table}[h!]
\centering
\caption{{Mechanical properties of different woven composites under load in X1 direction}}
\resizebox{0.8\textwidth}{!}{%
\begin{tabular}{cccccccc}
\hline
 $\alpha=\ang{45},\beta=\ang{90}$&
  \begin{tabular}[c]{@{}c@{}}CFRP \\ Woven  \end{tabular} &
  \begin{tabular}[c]{@{}c@{}}Artificial \\ Woven 1 \end{tabular} &
  \begin{tabular}[c]{@{}c@{}}Artificial \\ Woven 2 \end{tabular} &
  \begin{tabular}[c]{@{}c@{}}Artificial \\ Woven 3 \end{tabular} &
  \begin{tabular}[c]{@{}c@{}}Artificial \\ Woven 4 \end{tabular} &
  \begin{tabular}[c]{@{}c@{}}Artificial \\ Woven 5 \end{tabular} & 
  \begin{tabular}[c]{@{}c@{}}Artificial \\ Woven 6 \end{tabular} \\ \hline
Stiffness $K_c$ (N/mm) &
  3.15 &
  3.15 &
  3.15 &
  9.14 &
  9.14 &
  10.10 &
  3.37\\ 
 Poisson's ratio &
  -0.887 &
  -0.887 &
  -0.887 &
  -0.585 &
  -0.585 &
  -0.520 &
  -0.851\\ \hline
\end{tabular}%
}
\label{tab:mat2}
\end{table}

From the results in Table~\ref{tab:mat2}, we show that: the shear modulus has the most significant impact on the Miura-Ori structure’s compressive stiffness $K_c$ and Poisson's ratio (specifically NPR) compared to other parameters. On the other hand, compression modulus has the least influence, and the same trends are observed under load in the X3 direction. 

Moreover, to understand how individual shear modulus ($G_{12}$, $G_{23}$, $G_{13}$) controls the Miura-Ori structure's mechanical responses, we test different combinations of in-plane and out-of-plane shear modules and calculate the Increase Rate (IR) of $K_c$ and NPR for different shear modulus. Here IR is defined as the increase of targeting mechanical property based on the unit increase in shear modulus. We calculate the IR of the Miura-Ori structure with different values of $\alpha$ and $\beta$. Table~\ref{tab:increase_rate} shows the IR of some example models, where Model 1 has $\alpha=\ang{30}$, $\beta=\ang{90}$, Model 2 has $\alpha=\ang{45}$, $\beta=\ang{90}$ and Model 3 has $\alpha=\ang{75}$, $\beta=\ang{90}$. 

\begin{table}[h!]
\centering
\caption{Miura-Ori structure's mechanical responses Increase Rate (IR) under load in X1 direction}
\resizebox{0.8\textwidth}{!}{%
\begin{tabular}{ccccccc}
\hline
\multirow{2}{*}{Control material property} & \multicolumn{2}{c}{Model 1} & \multicolumn{2}{c}{Model 2} & \multicolumn{2}{c}{Model 3} \\  
                           & $K_c$ IR & NPR IR & $K_c$ IR & NPR IR  & $K_c$ IR & NPR IR    \\ \hline
In-plane shear modulus     & 0.13        & 5.26 & 0.25        & 22.90 & 1.52        & 137.20 \\ 
Out-of-plane shear modulus & 0.15        & 0.22 & 0.05        & 0.35  & 0.05        & 0.91   \\ \hline
\end{tabular}%
}
\label{tab:increase_rate}
\end{table}

where $K_c$ increase rate has the unit of $N/(mm*GPa)$ and NPR increase rate has the unit of $1/GPa$.

According to the results in Table~\ref{tab:increase_rate}, we conclude that: (1) the out-of-plane shear modulus is slightly dominant on the structure's stiffness ($K_c$) when $\alpha$ is small. As $\alpha$ increases, in-plane shear modulus rapidly becomes more significant for the structure's stiffness. (2) the in-plane shear modulus always plays a dominant role on NPR compared to the out-of-plane shear modulus. The same trend is also observed under loading in the X3 direction. Thus, we conclude that the in-plane shear modulus generally has the most significant role in controlling the Miura-Ori structure's mechanical response than the out-of-plane shear modulus.

\section{Conclusions}
This paper presents a detailed analysis of isotropic and directional Miura-Ori structures' mechanical responses using Finite Element Analysis. We investigated the Miura-Ori structure's optimal geometry for different materials and loading conditions. We proposed a regression model that can represent the relationship between the structure's geometric parameters ($\alpha$, $\beta$) and its mechanical response and determine the optimal shape analytically. At last, using the notion that flexibility in composite material properties could potentially be achieved using different weave patterns and yarn materials, we analyzed how individual material property is controlling the structure's mechanical responses. The relationship of material and geometric parameters to the Miura-Ori structure's mechanical responses explored in this paper and the regression model to analytically represent the Miura-Ori structure can be further used in the future to guide the Miura-Ori structure design and optimization.  

Key contributions to this paper are:

\begin{enumerate}
\item This is the first attempt to understand how material properties and geometric parameters influence the Miura-Ori structure, especially for directional Miura-Ori models (like CFRP composite). 

\item We analyze isotropic and directional Miura-Ori structures with different geometries. We conclude that composite materials within the Miura-Ori structure could improve the mechanical responses compared to the isotropic model. This leads to more flexibility in Miura-Ori structural design and application.  

\item We investigate the relationship between the geometric parameters and mechanical response (stiffness, negative Poisson’s ratio) of Miura-Ori structures with isotropic and directional materials. We determine that the optimal geometry is similar for different materials considered. 

\item To balance the optimization requirement between different mechanical responses, like stiffness and negative Poisson’s ratio, we present a regression analysis to analytically represent the mechanical responses as a function of geometric parameters. The regression models are further used to predict the Miura-Ori structure's mechanical responses analytically and deliver optimal design.

\item We show that shear modulus is dominant in controlling the Miura-Ori structure's stiffness and Poisson's ratio among different material properties. We demonstrate this by using woven composites and their flexibility in potentially changing the material properties within Miura-Ori structures.

\end{enumerate}

\section*{Acknowledgement}

The authors would like to acknowledge the financial support from the NSF CAREER award [\# 2046476] through the Mechanics of Materials and Structures (MOMS) Program for performing this research.

The authors would like to thank Hridyesh R. Tewani (University of Wisconsin - Madison) for the valuable discussions and inputs to the origami's manufacturing.

\section*{Data Availability}

All the Finite Element Analysis models considered in this research, corresponding files for generating plots shown in this paper, and the statistical analysis code can be found on google drive: \url{https://drive.google.com/drive/folders/1t6bmC_QHZssKYeoBERpcAuBTHto61olS?usp=sharing}.



{\footnotesize
\bibliographystyle{ieeetr}
\bibliography{sample}
}

\end{document}